\documentclass[bibyear]{aa} 

\usepackage{amsmath}
\usepackage{txfonts}
\usepackage{booktabs}
\usepackage{makecell}
\usepackage{hyperref}
    \hypersetup{
        colorlinks=true,
        linkcolor=blue,
        filecolor=magenta,      
        urlcolor=cyan,
        citecolor=blue
    }
\usepackage{color}
\usepackage{colortbl}
\usepackage[dvipsnames]{xcolor}
\usepackage{balance}
\usepackage{dcolumn}
\newcolumntype{d}[1]{D{.}{.}{#1}}
\newcommand\mc[1]{\multicolumn{1}{c}{#1}}
\usepackage{multicol}
\usepackage{multirow}
\newcommand\ms[1]{\vcenter{\hbox{$\scriptstyle #1$}}}

\begin{document} 

\authorrunning{X.\,L. Bacalla et al.}
\titlerunning{EDIBLES IV. Cosmic ray ionization rate from interstellar OH$^+$}
\title{The EDIBLES survey IV. Cosmic ray ionization rates in diffuse clouds from near-ultraviolet observations of interstellar OH$^+$}

    \author{Xavier L. Bacalla\inst{1}
        \and
        Harold Linnartz\inst{1}
        \and
        Nick L.\,J. Cox\inst{2,3}
        \and
        Jan Cami\inst{4,5}
        \and
        Evelyne Roueff\inst{6}
        \and
        Jonathan V. Smoker\inst{7}
        \and
        Amin~Farhang\inst{4,8}
        \and
        Jordy~Bouwman\inst{1}
        \and
        Dongfeng Zhao\inst{9}
        }
    
    \institute{
        Sackler Laboratory for Astrophysics, Leiden Observatory, P. O. Box 9513, NL-2300 RA Leiden, The Netherlands
        \and
        ACRI-ST, 260 Route du Pin Montard, 06904, Sophia Antipolis, France
        \and
        Anton Pannekoek Institute for Astronomy, University of Amsterdam, NL-1090 GE Amsterdam, The Netherlands
        \and
        Department of Physics and Astronomy, The University of Western Ontario, London, ON N6A 3K7, Canada
        \and
        SETI Institute, 189 Bernardo Avenue, Suite 100, Mountain View, CA 94043, USA
        \and
        Sorbonne Universit\'e, Observatoire de Paris, Universit\'e PSL,
        CNRS, LERMA, F-92190,  Meudon, France
        \and
        European Southern Observatory, Alonso de Cordova 3107, Vitacura, Santiago, Chile
        \and
        School of Astronomy, Institute for Research in Fundamental Sciences, 19395-5531 Tehran, Iran
        \and
        Hefei National Laboratory for Physical Sciences at the Microscale, Department of Chemical Physics, University of Science and Technology of China, Hefei, Anhui 230026, China
        }
    
    \date{Received 2018 / Accepted}
 
    \abstract{
    We report cosmic ray ionization rates towards ten reddened stars studied within the framework of the EDIBLES (ESO Diffuse Interstellar Bands Large Exploration Survey) program, using the VLT-UVES. For each sightline, between 2 and 10 individual rotational lines of OH$^+$ have been detected in its (0,0) and (1,0) $A^3\Pi-X^3\Sigma^-$ electronic band system. This allows constraining of OH$^+$ column densities towards different objects. Results are also presented for 28 additional sightlines for which only one or rather weak signals are found. An analysis of these data makes it possible to derive the primary cosmic ray ionization rate $\zeta_p$ in the targeted diffuse interstellar clouds. For the ten selected targets, we obtain a range of values for $\zeta_p$ equal to $(3.9-16.4) \times 10^{-16}~\mathrm{s}^{-1}$. These values are higher than the numbers derived in previous detections of interstellar OH$^+$ in the far-infrared / sub-millimeter-wave regions and in other near-ultraviolet studies. This difference is a result of using new OH$^+$ oscillator strength values and a more complete picture of all relevant OH$^+$ formation and destruction routes (including the effect of proton recombinations on PAHs), and the relatively high $N$(OH$^+$) seen toward those ten targets.
    } 
        

\maketitle

\section{Introduction}

    The hydroxyl cation, OH$^+$, is an important reactive intermediate in the gas phase formation of water in the diffuse interstellar medium (ISM) (van Dishoeck et al.~\cite{vanDishoeck+2013}), where ion-neutral molecule reactions are found to dominate (van Dishoeck $\&$ Black~\cite{vanDishoeck&Black1986}; Le Petit et al.~\cite{LePetit+2004}).  The formation mechanism of this ion involves the cosmic ray ionization of atomic or molecular hydrogen, followed by hydrogenation and oxygenation (Federman et al.~\cite{Federman+1996}; Hollenbach et al.~\cite{Hollenbach+2012}). Thus, apart from playing a role in interstellar water chemistry, OH$^+$ can also be used as a probe of the primary\footnote{The ``primary'' cosmic ray ionization rate $\zeta_p$ denotes the rate of ionization of atomic H that is solely caused by primary cosmic rays that have not interacted with the ISM to produce secondary particles.} cosmic ray ionization rate $\zeta_p$ in these dilute regions of molecular gas (Hollenbach et al.~\cite{Hollenbach+2012}; Porras et al.~\cite{Porras+2014}; Indriolo et al.~\cite{Indriolo+2015}).
    
    Rodebush $\&$ Wahl (\cite{Rodebush&Wahl1933}) first observed spectral lines of the OH$^+$ molecule in the laboratory and the recorded transitions were subsequently assigned by Loomis $\&$ Brandt (\cite{Loomis&Brandt1936}) to the $A^3\Pi-X^3\Sigma^-$ electronic band system in the near-UV. de Almeida $\&$ Singh (\cite{deAlmeida&Singh1981}) calculated the transition probabilities and oscillator strengths for these bands and suggested a number of wavelength positions where interstellar OH$^+$ is likely to manifest itself. de Almeida (\cite{deAlmeida1990}) also provided values for the rotational hyperfine transitions in the fundamental electronic state at sub-millimeter wavelengths. In searching for the 909~GHz (0.33~mm) transition of OH$^+$, Wyrowski et al.~(\cite{Wyrowski+2010}) detected this ion for the first time in space using the APEX telescope directed at Sagittarius B2(M). The 972~GHz (0.31~mm) transition was observed in a couple of bright continuum sources (in W31C by Gerin et al.~\cite{Gerin+2010} and in W49N by Neufeld et al.~\cite{Neufeld+2010}) through the Heterodyne Instrument for the Far-Infrared (HIFI) aboard the Herschel Space Observatory. Kre\l{}owski et al.~(\cite{Krelowski+2010}) detected a weak line at 3583.769~\AA\ in the spectra of a sample of interstellar sightlines obtained using the Ultraviolet and Visible Echelle Spectrometer (UVES) of the Very Large Telescope (VLT) that is due to an isolated rotational transition in the $A^3\Pi-X^3\Sigma^-$ electronic origin band system of OH$^+$. Porras et al.~(\cite{Porras+2014}) observed this same near-UV transition in a few other sightlines, and used it to estimate the value of $\zeta_p$ as was done in other work using sub-mm transitions (e.g., by Hollenbach et al.~(\cite{Hollenbach+2012})).
 
     \begin{table*}
        \caption{\label{tab:targets}EDIBLES targets with measurable OH$^+$ absorption.}
        \centering
        \begin{tabular}{lcccccccc}
        \toprule
        \midrule
        Identifier & \makecell{Galactic \\ Coordinates} & Spectral type & \makecell{$A_V$ \\ {[mag.]}} & \makecell{$E_{B-V}$ \\ {[mag.]}} & \makecell{$N$(H~\textsc{i}) \\ $\times10^{21}$~{[cm$^{-2}$]}} & \makecell{$N$(H\textsubscript{2}) \\ $\times10^{21}$~{[cm$^{-2}$]}} & \makecell{$N$(H\textsubscript{tot}) \\ $\times10^{21}$~{[cm$^{-2}$]}} & \makecell{$f_\mathrm{H\textsubscript{2}}$}\\
        \midrule
        HD~37367	& G179.0$-$01.0 &B2 IV-V	        &1.49	&0.37		&1.5	    &0.34	    &2.2 & 0.31\\ 
        HD~41117	& G189.6$-$00.8 &B2 Ia	            &1.25	&0.41		&2.5\tablefootmark{\textcolor{blue}{a}}	    &0.49	    &3.5 & 0.28\\
        HD~75860	& G264.1$+$00.2 &BC2 Iab            &3.10	&0.87		&$\cdots$	&$\cdots$	&$\cdots$&$\cdots$\\
        HD~79186	& G267.3$+$02.2 &B5 Ia	            &1.28	&0.28		&1.5	    &0.52	    &2.6 & 0.41\\
        HD~80558	& G273.0$-$01.4 &B6 Ia	            &2.01	&0.57		&$\cdots$	&$\cdots$	&$\cdots$&$\cdots$\\
        HD~114886	& G305.5$-$00.8 &O9 III+O9.5 III	&0.84	&0.28		&2.2	    &0.17	    &2.5 & 0.13\\
        HD~185418	& G053.6$-$02.1 &B0.5 V	            &1.27	&0.42		&1.6	    &0.51	    &2.6 & 0.39\\
        HD~185859	& G056.6$-$01.0 &B0.5 Ia            &1.64	&0.56		&1.7	    &$\cdots$	&$\geq$~1.7&$\cdots$\\ 
        HD~186745	& G060.2$-$00.2 &B8 Ia	            &2.98	&0.88		&$\cdots$	&$\cdots$	&$\cdots$&$\cdots$\\
        HD~186841	& G060.4$-$00.2 &B0.5 I	            &3.01	&0.95		&$\cdots$	&$\cdots$	&$\cdots$&$\cdots$\\
        \bottomrule
        \end{tabular}
        \tablefoot{The $A_V$-values are taken from Valencic et al.~(\cite{Valencic+2004}) and Wegner~(\cite{Wegner2003}), while the $E_{B-V}$-values are calculated by Cox et al.~(\cite{Cox+2017}). $N$(H~\textsc{i}) and $N$(H$_2$) data are taken from Jenkins~(\cite{Jenkins2009}) which are obtained through vacuum-UV absorption observations. The total hydrogen column density is calculated as $N$(H\textsubscript{tot}) = $N$(H~\textsc{i}) + $2 N$(H$_2$) while the molecular hydrogen fraction is calculated as $f_\mathrm{H\textsubscript{2}}$ = $2 N$(H$_2$)~/~$N$(H\textsubscript{tot}).\\
        \tablefootmark{a}{Diplas $\&$ Savage~\cite{Diplas&Savage1994}.}}
    \end{table*}
    
    Recently, Zhao et al.~(\cite{Zhao+2015}) detected (in four different sightlines) up to six of the near-UV OH$^+$ transitions initially provided by Merer et al.~(\cite{Merer+1975}), including two new (hitherto unidentified) interstellar features reported by Bhatt $\&$ Cami~(\cite{Bhatt&Cami2015}) in the same year. The detection of more than one transition makes it possible to derive a better constrained value for the OH$^+$ column density than that based on the detection of one transition only. The Zhao et al~(\cite{Zhao+2015}) results showed good agreement with previous measurements of $\zeta_p$ that were based not only on detections of OH$^+$, but also on detections of other cosmic ray ionization tracers in diffuse clouds like H$_2$O$^+$, H$_3$O$^+$, H$_3^+$, and ArH$^+$ (Neufeld et al.~\cite{Neufeld+2010}; Indriolo et al.~\cite{Indriolo+2012},~\cite{Indriolo+2015}; Le Petit et al.~\cite{LePetit+2004}; Indriolo et al.~\cite{Indriolo+2007}; Indriolo $\&$ McCall~\cite{Indriolo&McCall2012}; Neufeld $\&$ Wolfire~\cite{Neufeld&Wolfire2017}). More precise wavelengths and updated line oscillator strengths have also been provided recently through the updated spectral analyses and modelling efforts by Hodges \& Bernath (\cite{Hodges&Bernath2017}) and by Hodges et al. (\cite{Hodges+2018}). All of these previous work, and with more OH$^+$ lines now detected in the ISM, enable us to infer cosmic ray ionization rates in different sightlines where we have specifically chosen to characterize and know their physical properties as accurately as possible. With the vast spectral database and dedicated target characterization provided for by the ESO Diffuse Interstellar Bands Large Exploration Survey (EDIBLES) (Sec.~\ref{sec:Observations}), we report in this contribution detections of interstellar OH$^+$ via ground-based near-UV observations as a complement to sub-mm-wave observations such as with \emph{Herschel} (e.g., Gerin et al.~\cite{Gerin+2010}; Neufeld et al.~\cite{Neufeld+2010}) that require telluric-free conditions to record the pure rotational transitions of the OH$^+$ molecule. Since the OH$^+$ spectra are recorded as part of the DIB survey, this work also holds much potential in providing insight into the nature of the DIBs, as relevant physical parameters such as the cosmic ray ionization rate -- characterizing local conditions -- are needed to help further constrain the carriers of these enigmatic absorption features (Herbig~\cite{Herbig1995}; Cami $\&$ Cox~\cite{Cami&Cox2013}).

\section{Observations and Data Processing}\label{sec:Observations}

    The data used in this work were recorded within the framework of EDIBLES, or the ESO Diffuse Interstellar Bands Large Exploration Survey, which is a large (250+ hr) filler program (ESO ID 194.C-0833, PI. N.L.J. Cox) using the VLT-UVES in Paranal, Chile. Details are available from Cox et al.~(\cite{Cox+2017}). To briefly summarize, in this survey four standard configurations of UVES (Dekker et al.~\cite{Dekker+2000}) are employed, with wavelength settings centered at 3460, 4370, 5640, and 8600~\AA, covering from about 3042 to 10\,420~\AA, and with a spectral resolution of $\sim$~70\,000 in the blue ($<$~4800~\AA). A total of 114 unique sightlines are targeted and, as of May 2018, around 80 percent have been observed. For our particular application, we use spectra from the 346-nm setting (3042--3872~\AA) to look for and analyze the electronic transitions of OH$^+$, as well as the 437-nm setting (3752--4988~\AA) comprising the potassium (K~\textsc{i}) doublet that is used for estimating total hydrogen column densities (Sect.~3.2) in the different sightlines. All spectra presented here have been processed using standard and custom data reduction protocols (wavelength calibration, flat fielding, echelle order merging, etc.; see Cox et al.~(\cite{Cox+2017}) for details) and quality control by the EDIBLES team. 
        
    We searched the 93 available EDIBLES sightlines for the spectroscopic signature of OH$^+$, and clearly detected more than one transition in 10 targets. These are listed in Table~\ref{tab:targets} together with their respective galactic coordinates, spectral type, visual extinction $A_{V}$, and reddening $E_{B-V}$. Where available, the column densities for both atomic hydrogen (H or H~\textsc{i}) and molecular hydrogen (H$_2$) are also provided, as derived from measurements of L$\alpha$ and Lyman band absorptions, respectively (Diplas $\&$ Savage~\cite{Diplas&Savage1994}; Jenkins~\cite{Jenkins2009}). Also listed are the calculated column density of the total hydrogen atoms $N$(H\textsubscript{tot}) and the molecular hydrogen fraction $f_{\mathrm{H}_2}$ for each of the sightlines. Besides the ten selected targets with strongest detections, OH$^+$ was also seen along 28 other lines-of-sight. Here, signals were quite weak and typically limited to one transition. In principle, as will be shown, it is also possible to derive the cosmic ray ionization rate for these targets, but the resulting values are obviously much less accurate. (See Appendix~\ref{asec:moretargets} for an overview.)

\section{OH$^+$ as a probe of the cosmic ray ionization rate}\label{sec:CRIR}

\begin{figure}[]
        \centering
        \includegraphics[width=\hsize]{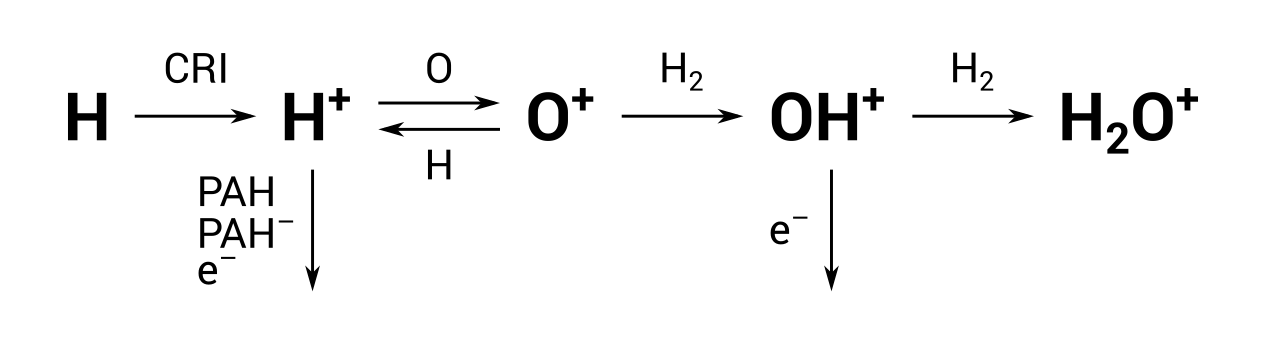}
        \caption{\label{fig:reaction}Ion-neutral chemistry of OH$^+$ through the atomic H reaction pathway (van Dishoeck \& Black~\cite{vanDishoeck&Black1986}; Hollenbach et al.~\cite{Hollenbach+2012}).}
    \end{figure}

    Cosmic rays are high-energy particles (mostly comprised of protons and helium nuclei) that originate from different astrophysical processes. They are ubiquitous across our galaxy and are one of the main drivers of ionization and chemistry in the ISM (Dalgarno~\cite{Dalgarno2006}). One of the important chemical processes that is influenced by cosmic ray ionization is the gas-phase formation scheme of water, in which the OH$^+$ molecule acts as a reactive intermediate. The dominant reaction pathway leading to the formation of OH$^+$ in diffuse clouds is depicted in Fig.~\ref{fig:reaction}.
    
    This reaction is initiated by the cosmic ray ionization (CRI) of atomic hydrogen, followed by charge exchange between H$^+$ and O --- producing O$^+$ which then reacts with H$_2$ to form OH$^+$. The H$^+$, however, can also react with electrons and with neutral and negatively charged PAHs. Through this reaction scheme, we can derive the rate of ionization due to cosmic rays by quantifying the abundance of OH$^+$ in these diffuse molecular environments. In dense molecular clouds, another important OH$^+$ formation pathway is through the ionization of molecular hydrogen, but we neglect this contribution in our formulation since we assume here that the medium is very diffuse while OH$^+$ is forming (Hollenbach et al.~\cite{Hollenbach+2012}).
    
    \begin{table}[h!]
        \caption{\label{tab:reaction}Reaction channels and rate coefficients.}
        \centering
        \resizebox{\columnwidth}{!} {%
        \begin{tabular}{lll}
        \toprule
        \midrule
        Reaction &  \multicolumn{2}{l}{Rate coefficient [cm$^{3}$\,s$^{-1}$]} \\
        \midrule
        
        Charge-transfer:\\
        \quad $\mathrm{H}^+ + \mathrm{O} \rightarrow \mathrm{O}^+ + \mathrm{H} $ &  $k_1$ & $4.0\times10^{-10}\mathrm{exp}(-227/T)$\tablefootmark{\textcolor{blue}{a}} \\
        \quad $\mathrm{O}^+ + \mathrm{H} \rightarrow \mathrm{H}^+ + \mathrm{O} $ &  $k_2$ & $4.0\times10^{-10}$\tablefootmark{\textcolor{blue}{a}} \\
        
        Ion-molecule reaction:\\
        \quad $\mathrm{O}^+ + \mathrm{H}_2 \rightarrow \mathrm{OH}^+ + \mathrm{H} $ &  $k_3$ & $1.7\times10^{-9}$ \\
        \quad $\mathrm{OH}^+ + \mathrm{H}_2 \rightarrow \mathrm{H}_2\mathrm{O}^+ + \mathrm{H} $ &  $k_4$ & $1.0\times10^{-9}$ \\
        
        H$^+$ and PAH$^{(-)}$ recombination:\\
        \quad $\mathrm{H}^+ + \mathrm{PAH} \rightarrow \mathrm{PAH}^+ + \mathrm{H} $ &  $\alpha$ & $7.0\times10^{-8}\Phi_\mathrm{PAH}$ \\
        \quad $\mathrm{H}^+ + \mathrm{PAH}^- \rightarrow \mathrm{PAH} + \mathrm{H} $ &  $\alpha^-$ & $8.1\times10^{-7}\Phi_\mathrm{PAH}\cdot(T/300)^{-0.50}$ \\
        
        Radiative recombination:\\
        \quad $\mathrm{H}^+ + \mathrm{e}^- \rightarrow \mathrm{H} + h\nu$ &  $\beta_\mathrm{H^+}$ & $3.5\times10^{-12}\cdot (T/300)^{-0.75}$ \\
        
        Dissociative recombination:\\
        \quad $\mathrm{OH}^+ + \mathrm{e}^- \rightarrow \mathrm{O} + \mathrm{H} $ &  $\beta_\mathrm{OH^+}$ & $3.8\times10^{-8}\cdot (T/300)^{-0.50}$ \\
        
        \midrule
        Cosmic ray ionization:\\
        \quad $\mathrm{H} + \mathrm{CR} \rightarrow \mathrm{H}^+ + \mathrm{e}^-$ &  $\zeta_\mathrm{H}$ & $\equiv 1.5 \zeta_p$ [s$^{-1}$]\tablefootmark{\textcolor{blue}{b}}
        \\
                
        \bottomrule
        \end{tabular}
        }
        \tablefoot{Rate coefficients are based on Hollenbach et al.~(\cite{Hollenbach+2012}) and references therein.\\
        \tablefootmark{a}Chambaud et al.~\cite{Chambaud+1980}.\\
        \tablefootmark{b}Glassgold \& Langer~\cite{Glassgold&Langer1974}.
        }
    \end{table}
        
    In order to establish a relation between CRI and OH$^+$, we must account for all formation and destruction routes that lead to OH$^+$, as well as for the intermediate species O$^+$ and H$^+$. These reaction channels are listed in Table~\ref{tab:reaction} with their respective reaction coefficients. With these, we build our formulation starting from three rate equations:
    
    \begin{align*}
        \begin{split}
        \frac{d\,\mathrm{OH}^+}{dt} &= k_3[\mathrm{O}^+][\mathrm{H}_2] - k_4[\mathrm{OH}^+][\mathrm{H}_2] - \beta_{\mathrm{OH}^+}[\mathrm{OH}^+][\mathrm{e}^-],
        \\
        \frac{d\,\mathrm{O}^+}{dt} &= k_1[\mathrm{H}^+][\mathrm{O}] - k_2[\mathrm{O}^+][\mathrm{H}] - k_3[\mathrm{O}^+][\mathrm{H}_2],\ \mathrm{and}
        \\
        \frac{d\,\mathrm{H}^+}{dt} &= \zeta_\mathrm{H}[\mathrm{H}] - \beta_{\mathrm{H}^+}[\mathrm{H}^+][\mathrm{e}^-] - k_1[\mathrm{H}^+][\mathrm{O}]
        \\
        &- \alpha[\mathrm{PAH}][\mathrm{H}^+] - \alpha^-[\mathrm{PAH}^-][\mathrm{H}^+],
        \end{split}
    \end{align*}
    
    \noindent and assuming that the rate of change in the density of atomic H can be neglected as these reactions occur. Setting each one to zero (for the steady-state condition) and combining all the terms gives
    
    \begin{align*}
        \begin{split}
        [\mathrm{OH}^+] &= \frac{ k_3[\mathrm{H}_2]k_1[\mathrm{O}] } { \big\{k_4[\mathrm{H}_2] + \beta_{\mathrm{OH}^+}[\mathrm{e}^-]\big\} \big\{k_2[\mathrm{H}] + k_3[\mathrm{H}_2]\big\} }~\times
        \\
        &\quad\,\frac{ \zeta_\mathrm{H}[\mathrm{H}] } { \big\{\beta_{\mathrm{H}^+}[\mathrm{e}^-] + k_1[\mathrm{O}] + \alpha[\mathrm{PAH}] + \alpha^-[\mathrm{PAH}^-]\big\} }.
        \end{split}
    \end{align*}
    
    \noindent This equation of densities can be further expressed in terms of the fractional abundance $x$ of species X, with respect to the total number of hydrogen atoms per unit volume $n_{\textrm{H}_{\textrm{tot}}}$ ($= [\textrm{H}] + 2[\textrm{H}_2]$) [cm$^{-3}$], that is, $x(\mathrm{X}) = [\mathrm{X}] / n_{\textrm{H}_{\textrm{tot}}}$. This then results into
    
    \begin{align*}
        \begin{split}
        x(\mathrm{OH}^+) &= \frac{ k_3x(\mathrm{H}_2)k_1x(\mathrm{O}) } { \big\{k_4x(\mathrm{H}_2) + \beta_{\mathrm{OH}^+}x_{\mathrm{e}^-}\big\} \big\{k_2x(\mathrm{H}) + k_3x(\mathrm{H}_2)\big\} }~\times
        \\
        &\quad\,\frac{ 1.5\zeta_px(\mathrm{H}) } { n_{\textrm{H}_{\textrm{tot}}} \big\{\beta_{\mathrm{H}^+} x_{\mathrm{e}^-} + k_1x(\mathrm{O})  + \alpha x(\mathrm{PAH})  + \alpha^- x(\mathrm{PAH}^-) \big\} }.
        \end{split}
    \end{align*}
    
    \noindent Since OH$^+$ is formed in a region where the molecular fraction is small, we can take $x$(H) = 10$x$(H$_2$) as with Porras et al.~(\cite{Porras+2014}) (see also Sec.~\ref{sec:Discussion}, paragraph 6). Combining this with $1 = x(\textrm{H}) + 2x(\textrm{H}_2)$ from above gives $x$(H)~=~0.833 and $x$(H$_2$)~=~0.083. For the rate coefficients [cm$^{3}$\,s$^{-1}$], assuming a kinetic temperature of $T = 100$~K for diffuse clouds, we take these values:
    $k_1 = 4.1 \times 10^{-11}$;
    $k_2 = 4.0 \times 10^{-10}$;
    $k_3 = 1.7 \times 10^{-9}$;
    $k_4 = 1.0 \times 10^{-9}$;
    $\beta_{\mathrm{H}^+} = 8.0 \times 10^{-12}$;
    $\beta_{\mathrm{OH}^+} = 6.6 \times 10^{-8}$.
    For the PAH$^{(-)}$ interactions, we adopt the scaling factor
    $\Phi_{\mathrm{PAH}} = 0.5$
    that takes care of the uncertainties in the PAH sizes and abundances (Wolfire et al.~\cite{Wolfire+2003}), yielding
    $\alpha = 3.5 \times 10^{-8}$~cm$^{3}$\,s$^{-1}$
    and
    $\alpha^- = 7.0 \times 10^{-7}$~cm$^{3}$\,s$^{-1}$.
    Then we take the total hydrogen density as
    $n_{\textrm{H}_{\textrm{tot}}} = 100$~cm$^{-3}$,
    the fractional ionization as
    $x_{\mathrm{e}^-} = 2 \times 10^{-4}$,
    and the fractional abundance of O as
    $x(\mathrm{O}) = 3 \times 10^{-4}$.
    Finally, we adopt the fractional abunances of PAHs and PAH anions as
    $x(\mathrm{PAH}) = 1.85 \times 10^{-7}$
    and
    $x(\mathrm{PAH^-}) = 1.5 \times 10^{-8}$,
    respectively (Hollenbach et al.~\cite{Hollenbach+2012}). Substituting all values gives us the expression for $\zeta_p$~[s$^{-1}$] in terms of the column densities of OH$^+$ and of the total hydrogen H\textsubscript{tot}:
    
    \begin{equation}\label{eq:CRIR}
    \zeta_p \approx 6.5 \times 10^{-8} \cdot \frac{N(\mathrm{OH}^+)}{N(\mathrm{H_{tot}})}.
    \end{equation}
    
    \noindent With this equation it is possible to calculate the primary cosmic ray ionization rate. This requires that for the individual sightlines, $N$(OH$^+$) is determined and that for $N$(H\textsubscript{tot}) the corresponding values are taken from Table~\ref{tab:targets} where $N$(H\textsubscript{tot}) = $N$(H~\textsc{i}) + $2 N$(H$_2$) or, alternatively, derived using the methods described in Sec.~\ref{sec:reddening&extinction} and \ref{sec:KI}.
    
    We want to stress that care is needed to compare the values that will be presented in the next section with those reported in the (recent) past. Eq.~\ref{eq:CRIR} has a different prefactor than used before because of incorporating different assumptions for the O$^+$ + H charge transfer rate coefficient ($4\times10^{-10}$~cm$^{3}$~s$^{-1}$ instead of $7\times10^{-10}$~cm$^{3}$~s$^{-1}$ which is based on Chambaud et al.~(\cite{Chambaud+1980}) and Stancil et al.~(\cite{Stancil+1999})) as well as introducing other sinks for H$^+$ ions. The temperature dependence of the prefactor is also already quite evident in some of the rate coefficients listed in Table~\ref{tab:reaction}, and this has a considerable effect on its resulting value. This will be discussed more in detail in Sec.~\ref{sec:Discussion}.

\section{Results}

    \subsection{Equivalent widths and column densities of OH$^+$}\label{ssec:EW}
    
\begin{figure}
        \centering
        \includegraphics[width=\hsize]{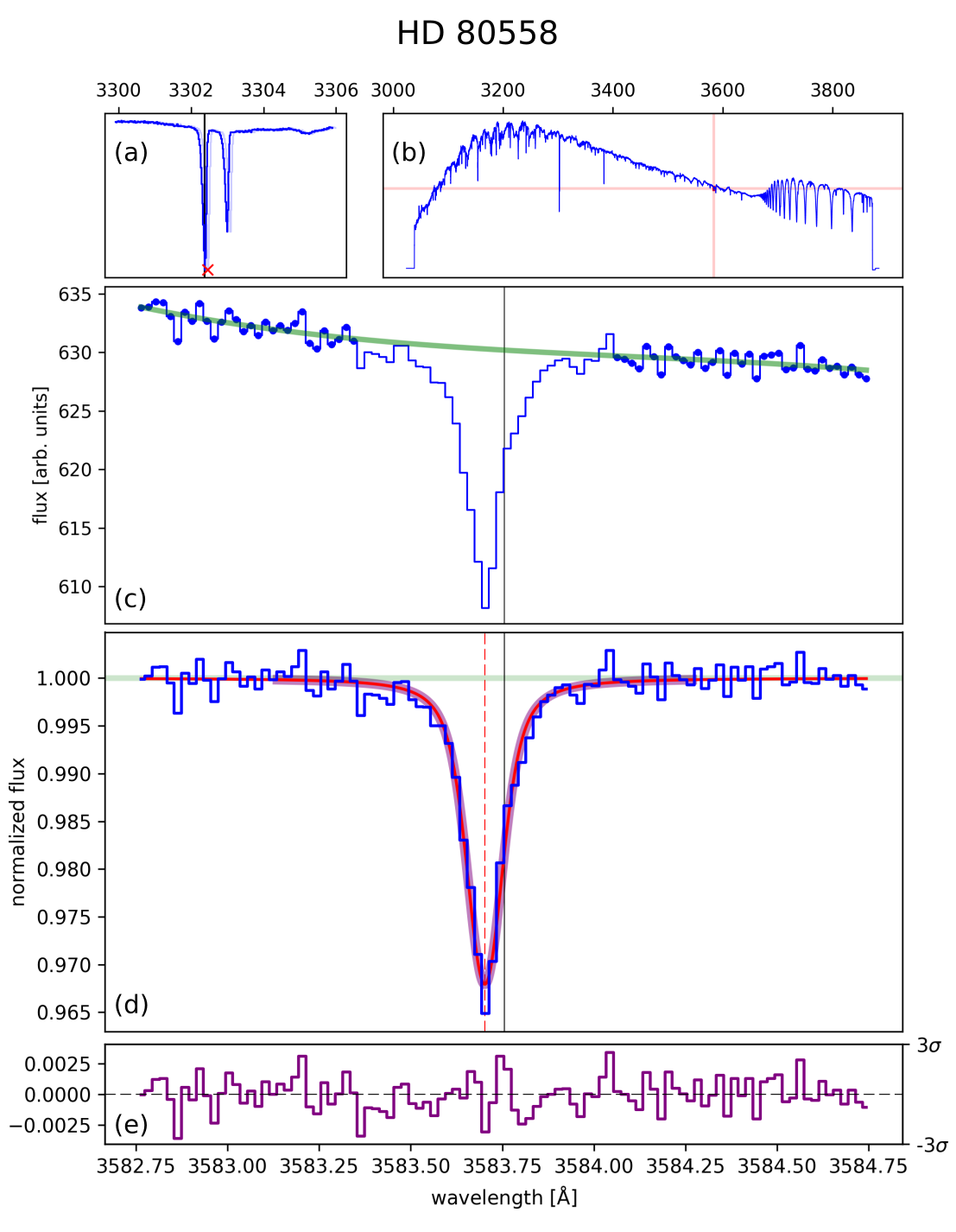}
        \caption{\label{fig:Voigt}Analysis of one OH$^+$ line (Line~1) for HD~80558. Description for each panel (a--e) can be found in the text.}
    \end{figure}

    The OH$^+$ electronic transitions listed in Table~\ref{tab:OH+lines} were searched for in the reduced spectra from the 346-nm setting (3042--3872~\AA). The procedure of analyzing the spectrum (developed in Python) is visualized in Fig.~\ref{fig:Voigt} for a selected OH$^+$ transition in one of the chosen stellar targets. Following heliocentric correction, the spectrum is shifted to the rest frame of one of the components of interstellar sodium (Na~\textsc{i} UV at 3302.3686~\AA\ in air (Kramida et al.~\cite{Kramida+2018}); panel a in the figure, marked with a red $\times$). Shown in panel b is a plot where the OH$^+$ line to be analyzed (the 3584~\AA\ transition) is located (indicated by the red crosshair) with respect to the full spectrum. A narrow wavelength range ($\sim$~1--2~\AA) is then selected, that includes the OH$^+$ line, after which data points are chosen where a polynomial (up to the 3rd order) is fitted (panel c). The zero-point velocity set by the Na~\textsc{i} UV transition is indicated by the solid vertical gray line. The selected spectrum is then divided by this fitted continuum for normalization. A Voigt function is fitted to the normalized spectrum, with which the equivalent width $W_\lambda$ is obtained by integrating (via Simpson integration) the area under the interpolated Voigt fit. The line is integrated within $\pm$~17 times the obtained sigma (= gamma) parameter from the central wavelength of the line.\footnote{The $\pm17\sigma$ integration bound is obtained empirically and is equivalent in area to $\pm2\cdot$FWHM used for a Gaussian profile.} The fitted profile is shown in panel d with the center wavelength indicated by the broken vertical red line. The corresponding residuals are shown in panel e. The resulting equivalent width is listed in Table~\ref{tab:summary}, together with those derived for other transitions and other sightlines. To estimate the uncertainty in the equivalent width calculation, we use the method described in Appendix A in the work of Vos et al.~(\cite{Vos+2011}).

    \begin{table}
        \caption{\label{tab:OH+lines}Rotational transitions in the $A^3\Pi-X^3\Sigma^-$ electronic band system of OH$^+$.}
        \centering
        \begin{tabular}{ccclc}
        \toprule
        \midrule
        \multicolumn{2}{c}{Transition} & Label & \mc{$\lambda$\ [\AA]} & \mc{$f\times10^{-4}$} \\
        \midrule
        (0,0) & $^rR_{11}(0)$ & 1  & 3583.75574(16) & 5.27  \\ 
              & $^rQ_{21}(0)$ & 2  & 3572.65187(33)   & 3.12  \\ 
              & $^sR_{21}(0)$ & 3  & 3566.4458(11)  & 1.17  \\ 
              & $^rP_{31}(0)$ & 4  & 3565.34592(81)   & 1.28  \\ 
              & $^sQ_{31}(0)$ & 5  & 3559.8062(13)  & 0.87  \\ 
              & $^tR_{31}(0)$ & 6  & 3552.325(12)   & 0.05  \\ 
        \midrule
        (1,0) & $^rR_{11}(0)$ & 7  & 3346.95559(74)   & 3.52  \\
              & $^rQ_{21}(0)$ & 8  & 3337.3570(15)  & 2.06  \\
              & $^sR_{21}(0)$ & 9  & 3332.177(11)   & 0.82  \\
              & $^rP_{31}(0)$ & 10 & 3330.409(11)   & 0.85  \\
              & $^sQ_{31}(0)$ & 11 & 3326.369(11)   & 0.62  \\
              & $^tR_{31}(0)$ & 12 & 3319.967(11)   & 0.04  \\
        \bottomrule
        \end{tabular}
        \tablefoot{The wavelengths (in standard air) and the line oscillator strengths are taken from Hodges \& Bernath~(\cite{Hodges&Bernath2017}) and Hodges et al.~(\cite{Hodges+2018}). Numbers enclosed in parentheses denote the uncertainty of the last digits.
        }
    \end{table}
    
    \begin{figure*}
        \centering
        \includegraphics[width=\textwidth]{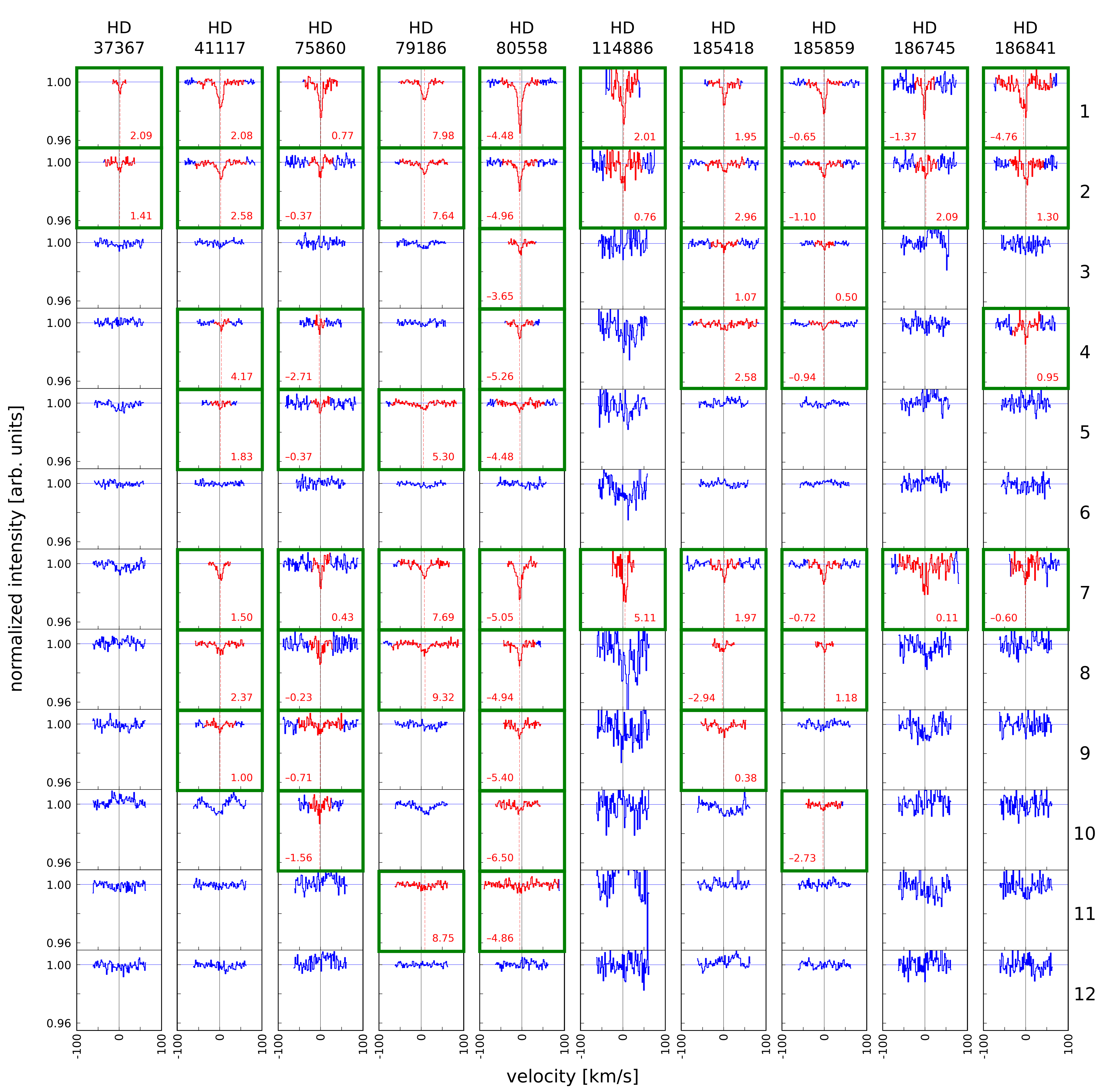}
        \caption{\label{fig:grid}The OH$^+$ absorption lines observed (enclosed in green boxes) for each of the 10 stellar targets. Every row of boxes is labeled on the right side according to the line labeling in Table~\ref{tab:OH+lines}. The solid vertical gray line in each box denotes the zero-point velocity set by the Na~\textsc{i} UV transition, while the broken vertical red line denotes the center of the fitted profile. The inset of numbers indicates the velocity difference in km\,s$^{-1}$ of the OH$^+$ line from the rest frame of Na~\textsc{i} UV. The red-highlighted trace in the spectrum is the region where the fitted Voigt profile is integrated. Note that the y-scaling is uniform to emphasize the relative strengths of each of the lines in one target and among the rest. The relative S/N can also be directly compared.
        }
    \end{figure*}

    The observed OH$^+$ lines for each of the 10 targets are compiled in Fig.~\ref{fig:grid}. As can be seen in the figure, between two and ten absorption lines are found in each target, with Lines 1, 2, and 7 being the most intense as expected from the magnitude of their oscillator strengths. The signal-to-noise ratio of the spectra in this wavelength range varies from 100 to 1200 per pixel (median value $\sim500$) which allows for many of the weaker lines to be observed. In some of the targets, the region where the OH$^+$ line is observed exhibits a background continuum which can be harder to fit using a low-order polynomial. This can be due to jumps in the spectrum after echelle order merging or blending absorption from other species -- stellar lines such as, e.g., Fe~\textsc{ii} at 3566.2~\AA, Ti~\textsc{ii} at 3332.1~\AA, or Mn~\textsc{ii} at 3330.8~\AA\ (Kurucz \& Bell~\cite{Kurucz&Bell1995}) -- which also explains why some of the weaker lines show up more than the stronger ones if they are by chance in a region with a better defined continuum. In these cases, care is taken to only include a small part of the spectrum (highlighted in red in the figure) in the equivalent width calculation.
        
    The column density $N(\mathrm{OH}^+)$ along a particular sightline is obtained by plotting the equivalent width [m\AA] of each absorption line against the product of the corresponding oscillator strength and the square of its wavelength [\AA] as defined by the following approximation valid for an optically-thin absorber (Spitzer~\cite{Spitzer1978}):

    \begin{equation}\label{eq:EW}
    W_\lambda = N(\mathrm{OH}^+) \cdot 8.853 \times 10^{-18} \cdot \lambda^2 f.
    \end{equation}

\begin{figure}
        \centering
        \includegraphics[width=\hsize]{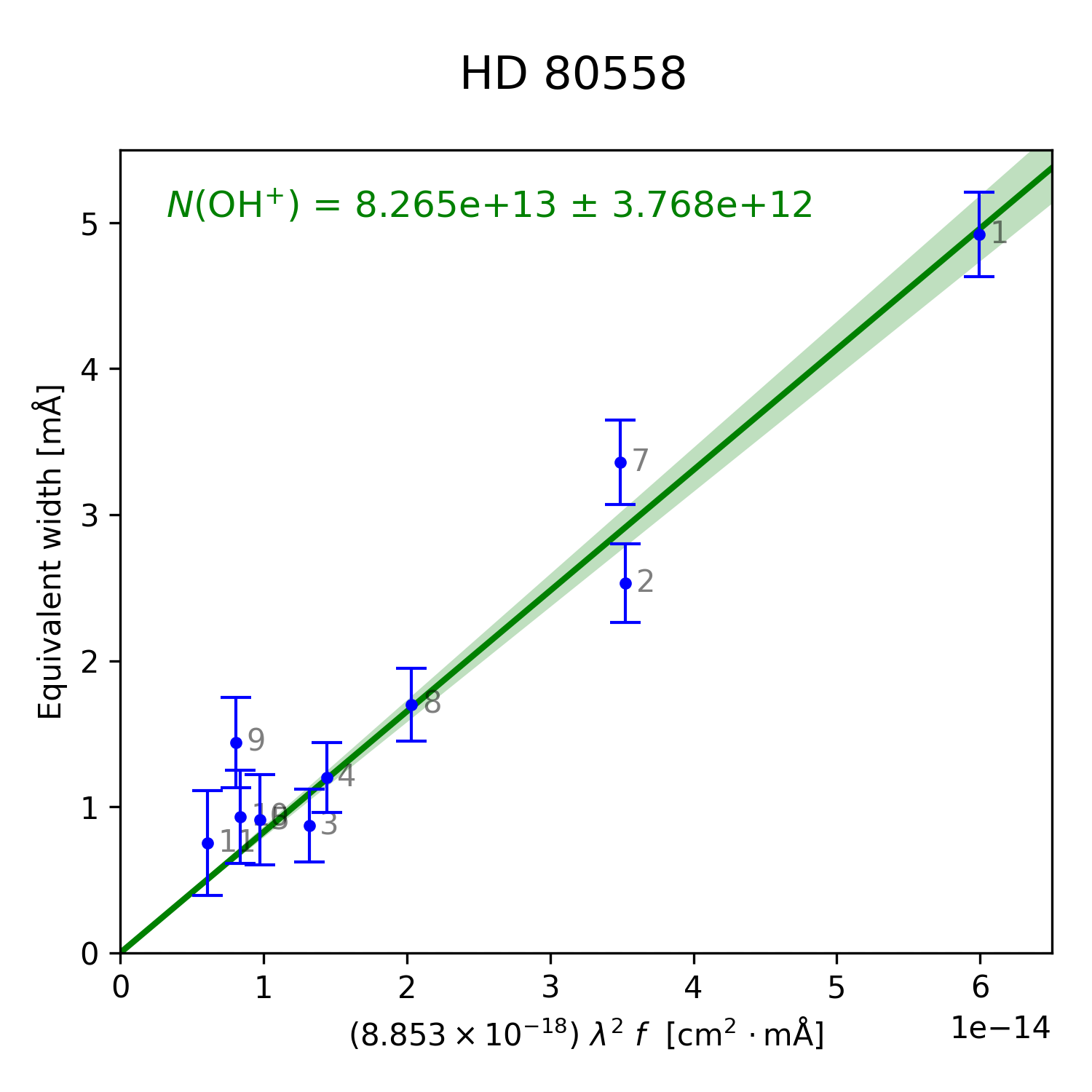}
        \caption{\label{fig:N(OH+)}The OH$^+$ column density (shown here for HD~80558) is derived from the slope of the line through points, plotting the equivalent width values as function of the transition wavelength and the oscillator strength, using Eq.~\ref{eq:EW}. The line through the points is a linear fit weighted to the uncertainty of $W_\lambda$ which yields a slope of $(8.3 \pm 0.4) \times 10^{13}$~cm$^{-2}$. Data points are marked according to the corresponding label for the OH$^+$ line (Table~\ref{tab:OH+lines}). (See Fig.~\ref{afig:N(OH+)} in Appendix~\ref{asec:N(OH+)} for the plots of the other targets.)
        }
    \end{figure}

    \noindent Eq.~\ref{eq:EW} can be used to fit a linear curve since all involved transitions originate from the same ground state level defined by the angular momentum quantum number $N = 0$. The column density is then obtained from the slope of this equation, as depicted in Fig.~\ref{fig:N(OH+)}. The linear fit is made to intercept the origin of the graph following the obvious assumption that no equivalent width can be measured if no absorption line exists. The derived column densities for each sightline are listed in Table~\ref{tab:summary}. With more than one OH$^+$ transition observed per sightline, this method allows for a better constraint in the resulting $N(\mathrm{OH}^+)$-value. As an added advantage, employing more than one transition also reduces the impact of having coincidental stellar contamination on any particular OH$^+$ line to the final measured $N$(OH$^+$)-value.
    
    As stated before, some 28 additional sightlines show (weak) OH$^+$ detections, typically comprising of only one transition. In principle, it is possible to derive values for $\zeta_p$ through Eq.~\ref{eq:EW}. These are listed in Appendix~\ref{asec:moretargets} as well. However, as the Voigt fitting is complicated as some of these lines have multiple components, and the $N$(OH$^+$)-values will only be derived from one OH$^+$ absorption line, we restrict our main conclusions to the ten targets as listed in Table~\ref{tab:targets}. For the non-detections of OH$^+$ for all available sightlines in EDIBLES, we have derived the 5$\sigma$ equivalent width upper limit for the strongest OH$^+$ line, which was estimated as $W_{\lambda(\mathrm{limit})} = 5\sigma\sqrt{N}(\Delta\lambda)$ (Jenkins et al.~\cite{Jenkins+1973}), with $\Delta\lambda$ as the width of each wavelength bin (binsize = 0.02~\AA), \textit{N} as the number of points included in the sampled $\lambda$3854 OH$^+$ line (\textit{N} = 28), and $\sigma$ as the reciprocal of the signal-to-noise ratio SNR around the line. An average value of 1.0~m\AA\ is obtained, corresponding to an upper limit column density $N$(OH$^+$)\textsubscript{(limit)} of $2.1\times10^{13}$~cm$^{-2}$ (with a range of $0.7-13.5\times10^{13}$~cm$^{-2}$).\balance
    
\subsection{Complementary methods for deriving N(H\textsubscript{tot})}\label{ssec:N(Htot)}

    Now that the $N$(OH$^+$)-values are known, the other quantity that is needed to derive $\zeta_p$ (Eq.~\ref{eq:CRIR}) is the total hydrogen column density $N$(H\textsubscript{tot}) along each of the sightlines. As was described shortly in Sec.~\ref{sec:CRIR}, we can directly obtain this from Table~\ref{tab:targets} with $N$(H\textsubscript{tot}) = $N$(H~\textsc{i}) + $2 N$(H$_2$). In the next sections we will describe additional ways that we have used for deriving $N$(H\textsubscript{tot}) based on interstellar reddening $E_{B-V}$ or potassium (K~\textsc{i}) absorption line measurements. The resulting values are summarized in Table~\ref{tab:summary}.

    \subsubsection{\emph{N}(H\textsubscript{tot}) from interstellar reddening \emph{E}\textsubscript{\emph{B}--\emph{V}}}
    \label{sec:reddening&extinction}
    
    $N$(H\textsubscript{tot}) is commonly estimated through interstellar reddening using the relation identified by Bohlin et al.~\cite{Bohlin+1978} for diffuse clouds: $N(\mathrm{H}_\mathrm{tot}) = 5.8 \times 10^{21} \cdot E_{B-V}$~cm$^{-2}$. As in the work done by Jenkins~(\cite{Jenkins2009}), these values and relations of $N$(H\textsubscript{tot}) with interstellar reddening were determined using L$\alpha$ absorption line measurements. Since the entire sightline is also considered in these measurements, the total number of hydrogen atoms associated with the production and destruction of OH$^+$ in the local interstellar cloud(s) may be overestimated. Nevertheless, these $N$(H\textsubscript{tot})-values can be easily derived, using the $E_{B-V}$-values listed in Table~\ref{tab:targets}.

\subsubsection{\emph{N}(H\textsubscript{tot}) from potassium absorption line measurements}\label{sec:KI}
    
    Another independent way of deriving $N$(H\textsubscript{tot}) is through an absorption line measurement of the interstellar K~\textsc{i} doublet at 4044.1422 and 4047.2132~\AA, with $f_{4044} = 5.69\times10^{-3}$ and $f_{4047} = 2.63\times10^{-3}$, respectively (Kramida et al.~\cite{Kramida+2018}). This approach is based on the empirical relation presented by Welty \& Hobbs~(\cite{Welty&Hobbs2001}), derived from high resolution spectral observations, that is,
    
    \begin{equation}\label{eq:W&H}
        \mathrm{log}\big[N(\mathrm{K~\textsc{i}})\big] = A + B \cdot \mathrm{log}\big[N(\mathrm{H_{tot}})\big].
    \end{equation}

    \noindent Here, the values chosen for coefficients $A$ and $B$ are $-26.30 \pm 1.09$ and $1.79 \pm 0.16$, respectively. These numbers were obtained when all stars in their sample were included in the analysis (Table~5 in Welty \& Hobbs~(\cite{Welty&Hobbs2001})). The column density $N$(K~\textsc{i}) is obtained via a similar manner as with $N$(OH$^+$), i.e., using the same fitting routine. The equivalent width of each of the two doublet components is then fitted with a line through the origin as defined by Eq.~\ref{eq:EW}. This linear fit can be used since the majority of the measured $W_\lambda$-values for the K~\textsc{i} doublet follows the ratio of their oscillator strengths, i.e., $W_{4044}:2W_{4047} \approx f_{4044}:2f_{4047}$, which indicates that the absorption lines are well within the thin-absorber approximation. The results of these calculations are summarized in Table~\ref{tab:summary}.
    
    \begin{table*}[]
        \centering
        \caption{Summary of measured and calculated values.}
        \label{tab:summary}
        \resizebox{\textwidth}{!}{%
        \begin{tabular}{@{}lcllcclccccccccccc@{}}
        \toprule
        \midrule
        \multirow{3}{*}{Identifier} &  & \multicolumn{3}{c}{OH$^+$} &  & \multicolumn{2}{c}{K$~\textsc{i}$} &  & \multicolumn{3}{c}{$N$(H$_\mathrm{tot}$) $\times 10^{21}$ {[}cm$^{-2}${]}} &  & \multicolumn{3}{c}{$\zeta_p$ $\times 10^{-16}$ {[}s$^{-1}${]}} \\
        \cline{3-5} \cline{7-8} \cline{10-12} \cline{14-16} \rule{0pt}{2.5ex}
         & & \multicolumn{2}{c}{$W_\lambda$ [m$\AA$]}  & \mc{$N$(OH$^+$)} &  & \multirow{2}{*}{$W_\lambda$ [m$\AA$]} & \mc{$N$(K$~\textsc{i}$)} &  & \multirow{2}{*}{Table 1}              & \multirow{2}{*}{$E_{B-V}$}           & \multirow{2}{*}{$N$(K$~\textsc{i}$)}           &  & \multirow{2}{*}{Table 1}         & \multirow{2}{*}{$E_{B-V}$}        & \multirow{2}{*}{$N$($K~\textsc{i}$)}       \\
         & & \mc{(0,0)}  & \mc{(1,0)} & \mc{$\times 10^{13}$ [cm$^{-2}$]} & & & \mc{$\times 10^{13}$ [cm$^{-2}$]} & & & & & & & & \\
        \midrule
        HD 37367                    &  & $\ms{1}$ 0.7(4)        & & 1.3(3)                                                                                &  & \mc{$\cdots$}                 & $\cdots$                                                                                       &  & 2.2                  & 2.15                & $\cdots$                      &  & 3.9(8)        & 3.9(8)         & $\cdots$                  \\
                                    &  & $\ms{2}$ 0.7(5)          &&                                                                                        &  &                          &                                                                                                &  &                      &                     &                               &  &                 &                  &                           \\
                                    &  &                          &                                                                                        &  &                          &                                                                                                &  &                      &                     &                               &  &                 &                  &                           \\
        HD 41117                    &  & $\ms{1}$ 3.3(3)          &$\ms{7}$ 1.8(3)& 4.9(6)                                                                                 &  & $\ms{1}$ 1.4(2)        & 0.15(2)                                                                                       &  & 3.5                  & 2.38                & 3.1(4)                       &  & 8.9(1.1)            & 13.2(1.6)            & 10.0(1.8)                     \\
                                    &  & $\ms{2}$ 2.2(3)          &$\ms{8}$ 1.2(3)&                                                                                        &  & $\ms{2}$ 0.5(2)        &                                                                                                &  &                      &                     &                               &  &                 &                  &                           \\
                                    &  & $\ms{4}$ 0.4(2)        &$\ms{9}$ 0.6(2)&                                                                                        &  &                          &                                                                                                &  &                      &                     &                               &  &                 &                  &                           \\
                                    &  & $\ms{5}$ 0.4(2)        &&                                                                                        &  &                          &                                                                                                &  &                      &                     &                               &  &                 &                  &                           \\
                                    &  &                          &                                                                                        &  &                          &                                                                                                &  &                      &                     &                               &  &                 &                  &                           \\
        HD 75860                    &  & $\ms{1}$ 3.2(7)& $\ms{7}$ 1.2(5)          & 3.9(5)                                                                                 &  & $\ms{1}$ 0.3(2)        & 0.05(1)                                                                                      &  & $\cdots$             & 5.05                & 1.6(4)                        &  & $\cdots$        & 5.0(7)           & 15.7(4.2)                     \\
                                    &  & $\ms{2}$ 0.9(6)& $\ms{8}$ 1.2(5)          &                                                                                        &  & $\ms{2}$ 0.3(2)         &                                                                                                &  &                      &                     &                               &  &                 &                  &                           \\
                                    &  & $\ms{4}$ 0.3(3)& $\ms{9}$ 0.9(8)        &                                                                                        &  &                          &                                                                                                &  &                      &                     &                               &  &                 &                  &                           \\
                                    &  & $\ms{5}$ 0.5(5)& $\ms{10}$ 0.5(3)          &                                                                                        &  &                          &                                                                                                &  &                      &                     &                               &  &                 &                  &                           \\
                                    &  &                          &                                                                                        &  &                          &                                                                                                &  &                      &                     &                               &  &                 &                  &                           \\
        HD 79186                    &  & $\ms{1}$ 2.6(3)& $\ms{7}$ 1.8(3)        & 4.8(4)                                                                                 &  & $\ms{1}$ 0.4(2)         & 0.05(1)                                                                                       &  & 2.6                  & 1.62                & 1.7(3)                       &  & 12.1(1.1)           & 19.0(1.8)            & 18.3(3.5)                     \\
                                    &  & $\ms{2}$ 1.6(3)& $\ms{8}$ 1.5(4)        &                                                                                        &  & $\ms{2}$ 0.3(2)         &                                                                                                &  &                      &                     &                               &  &                 &                  &                           \\
                                    &  & $\ms{5}$ 1.1(4)& $\ms{11}$ 0.9(4)          &                                                                                        &  &                          &                                                                                                &  &                      &                     &                               &  &                 &                  &                           \\
                                    &  &                          &                                                                                        &  &                          &                                                                                                &  &                      &                     &                               &  &                 &                  &                           \\
        HD 80558                    &  & $\ms{1}$ 4.9(3)& $\ms{7}$ 3.4(3)          & 8.3(4)                                                                                 &  & $\ms{1}$ 2.0(1)        & 0.19(7)                                                                                        &  & $\cdots$             & 3.31                & 3.6(1.3)                          &  & $\cdots$        & 16.3(7)          & 14.9(5.5)                     \\
                                    &  & $\ms{2}$ 2.5(3)& $\ms{8}$ 1.7(3)          &                                                                                        &  & $\ms{2}$ 0.3(1)         &                                                                                                &  &                      &                     &                               &  &                 &                  &                           \\
                                    &  & $\ms{3}$ 0.9(3)& $\ms{9}$ 1.4(3)        &                                                                                        &  &                          &                                                                                                &  &                      &                     &                               &  &                 &                  &                           \\
                                    &  & $\ms{4}$ 1.2(2)& $\ms{10}$ 0.9(3)        &                                                                                        &  &                          &                                                                                                &  &                      &                     &                               &  &                 &                  &                           \\
                                    &  & $\ms{5}$ 0.9(3)& $\ms{11}$ 0.8(4)          &                                                                                        &  &                          &                                                                                                &  &                      &                     &                               &  &                 &                  &                           \\
                                    &  &                          &                                                                                        &  &                          &                                                                                                &  &                      &                     &                               &  &                 &                  &                           \\
        HD 114886                   &  & $\ms{1}$ 3.3(1.4)& $\ms{7}$ 2.9(1.5)          & 6.1(8)                                                                                 &  & \mc{$\cdots$}                 & $\cdots$                                                                                       &  & 2.5                  & 1.62                & $\cdots$                      &  & 15.8(2.1)           & 24.6(3.3)            & $\cdots$                  \\
                                    &  & $\ms{2}$ 2.1(1.6)          &&                                                                                        &  &                          &                                                                                                &  &                      &                     &                               &  &                 &                  &                           \\
                                    &  &                          &                                                                                        &  &                          &                                                                                                &  &                      &                     &                               &  &                 &                  &                           \\
        HD 185418                   &  & $\ms{1}$ 1.9(5)& $\ms{7}$ 1.3(4)          & 3.3(3)                                                                                 &  & $\ms{1}$ 1.1(3)        & 0.132(3)                                                                                      &  & 2.6                  & 2.44                & 2.9(1)                      &  & 8.4(8)          & 8.9(8)           & 7.4(7)                      \\
                                    &  & $\ms{2}$ 0.9(5)& $\ms{8}$ 0.7(5)          &                                                                                        &  & $\ms{2}$ 0.5(3)        &                                                                                                &  &                      &                     &                               &  &                 &                  &                           \\
                                    &  & $\ms{3}$ 0.5(4)& $\ms{9}$ 1.4(7)          &                                                                                        &  &                          &                                                                                                &  &                      &                     &                               &  &                 &                  &                           \\
                                    &  & $\ms{4}$ 0.9(7)          &&                                                                                        &  &                          &                                                                                                &  &                      &                     &                               &  &                 &                  &                           \\
                                    &  &                          &                                                                                        &  &                          &                                                                                                &  &                      &                     &                               &  &                 &                  &                           \\
        HD 185859                   &  & $\ms{1}$ 3.0(4)& $\ms{7}$ 1.7(4)          & 4.3(4)                                                                                 &  & $\ms{1}$ 1.4(2)         & 0.175(1)                                                                                       &  & $\geq$ 1.7           & 3.25                & 3.40(1)                       &  & $\leq$ 16.4           & 8.6(8)    & 8.2(8)                    \\
                                    &  & $\ms{2}$ 1.5(4)& $\ms{8}$ 0.5(3)          &                                                                                        &  & $\ms{2}$ 0.7(2)         &                                                                                                &  &                      &                     &                               &  &                 &                  &                           \\
                                    &  & $\ms{3}$ 0.4(3)& $\ms{10}$ 0.5(4)        &                                                                                        &  &                          &                                                                                                &  &                      &                     &                               &  &                 &                  &                           \\
                                    &  & $\ms{4}$ 0.5(4)        &&                                                                                        &  &                          &                                                                                                &  &                      &                     &                               &  &                 &                  &                           \\
                                    &  &                          &                                                                                        &  &                          &                                                                                                &  &                      &                     &                               &  &                 &                  &                           \\
        HD 186745                   &  & $\ms{1}$ 1.9(7)& $\ms{7}$ 3.0(1.1)          & 3.4(8)                                                                                 &  & $\ms{1}$ 3.7(6)          & 0.38(7)                                                                                      &  & $\cdots$             & 5.10                & 5.3(1.0)                        &  & $\cdots$        & 4.3(1.0)           & 4.2(1.3)                    \\
                                    &  & $\ms{2}$ 0.9(7)          &&                                                                                        &  & $\ms{2}$ 1.2(4)        &                                                                                                &  &                      &                     &                               &  &                 &                  &                           \\
                                    &  &                          &                                                                                        &  &                          &                                                                                                &  &                      &                     &                               &  &                 &                  &                           \\
        HD 186841                   &  & $\ms{1}$ 4.1(1.2) & $\ms{7}$ 1.3(8)          & 5.3(9)                                                                                 &  & $\ms{1}$ 3.3(2)          & 0.395(1)                                                                                       &  & $\cdots$             & 5.51                & 5.37(1)                       &  & $\cdots$        & 6.3(1.1)             & 6.4(1.1)                      \\
                                    &  & $\ms{2}$ 1.8(1.0)          &&                                                                                        &  & $\ms{2}$ 1.5(3)          &                                                                                                &  &                      &                     &                               &  &                 &                  &                           \\
                                    &  & $\ms{4}$ 1.3(8)          &&                                                                                        &  &                          &                                                                                                &  &                      &                     &                               &  &                 &                  &                           \\
        \midrule
        \multicolumn{1}{c}{}        &  &                  &&  &    &                 &                     &  & \multicolumn{3}{r}{\textit{$\zeta_p$ weighted average:}}                   &  & 8.5(4)        & 8.5(3)         & 7.5(4) \\
        \bottomrule
        \end{tabular}%
        }
        \tablefoot{For the OH$^+$ equivalent width measurements in column 2, each of the values is preceded with a number labelling the absorption line as listed in Table~\ref{tab:OH+lines}; for the K~\textsc{i} $W_\lambda$-values, labels 1 and 2 represent the 4044~\AA\ and the 4047~\AA\ components, respectively. The total hydrogen column density $N$(H\textsubscript{tot}) is obtained via three ways: derived from Table~\ref{tab:targets} where $N$(H\textsubscript{tot}) = $N$(H~\textsc{i}) + $2 N$(H$_2$), from $E_{B-V}$, and from $N$(K~\textsc{i}) (columns 7--9). The corresponding cosmic ray ionization rates $\zeta_p$ calculated using each of these $N$(H\textsubscript{tot})-values are listed in columns 10--12. Numbers enclosed in parentheses denote the uncertainty of the last digit(s); e.g., 1.2(3) $\equiv$ 1.2~$\pm$~0.3, whereas 4.3(2.1) $\equiv$ 4.3~$\pm$~2.1.
        }
    \end{table*}

\section{Discussion}\label{sec:Discussion}
    
    The calculated cosmic ray ionization rates are summarized in Table~\ref{tab:summary}. When we compare the resulting $\zeta_p$-values using the different procedures of deriving $N$(H\textsubscript{tot}), we see a range of values around $(3.9-24.6) \times 10^{-16}~\mathrm{s}^{-1}$. In sightlines where information on both K~\textsc{i} and literature values for $N$(H\textsubscript{tot}) are available, some discrepancy is found between the calculated $\zeta_p$ using the two procedures, varying from a factor of 0.7 (HD~79186) to $\leq$~2.0 (HD~185859). On the other hand, it is interesting to find that the values derived using $E_{B-V}$ are not very different to those obtained through K~\textsc{i} measurements (within 20 percent), apart from the data of HD~75860, knowing that the corresponding $N$(H\textsubscript{tot})-values can be over- or underestimated. Overall, there exists a reasonable agreement for the different approaches discussed here which is also evident in the weighted averages of the $\zeta_p$-values.  However, when comparing with other work below, we instead quote $(3.9-16.4) \times 10^{-16}~\mathrm{s}^{-1}$, with a weighted average of $8.5(4) \times 10^{-16}~\mathrm{s}^{-1}$; this result only includes the $\zeta_p$-values derived using the $N$(H\textsubscript{tot}) from Table~\ref{tab:targets} since these come from more reliable and direct vacuum-UV measurements of $N$(H~\textsc{i}) and $N$(H$_2$). These $N$(H\textsubscript{tot})-values will serve as upper limits to the actual amount of hydrogen that is involved in the formation of OH$^+$.

    Before we can start comparing our results with other near-UV studies, it is important to note that Porras et al.~(\cite{Porras+2014}) and Zhao et al.~(\cite{Zhao+2015}) have adopted a similar formula (Eq.~\ref{eq:CRIR}) but with a prefactor ($\sim1.3~\times~10^{-8}$~s$^{-1}$) which is five times smaller, based on the rate equations provided by Federman et al.~(\cite{Federman+1996}). As discussed in Sec.~\ref{sec:CRIR}, in these studies the recombination of protons on PAHs was not taken into account. We have also left out the He$^+$ recombination rate found in their formulation after considering the reaction channels listed in Table~\ref{tab:reaction}.
    Apart from the new prefactor, it should also be noted that new OH$^+$ line oscillator strengths from Hodges et al.~(\cite{Hodges+2018}) were used for updating the results obtained from previous near-UV work (Porras et al.~\cite{Porras+2014}; Zhao et al.~\cite{Zhao+2015}) for a fully consistent comparison. The updated results are listed in Table~\ref{atab:literature} together with the original values previously reported.
    
    The adapted $\zeta_p$-values from Zhao et al.~(\cite{Zhao+2015}), with a range of $(6.6-11.1) \times 10^{-16}~\mathrm{s}^{-1}$ and a weighted average of $8.6(2) \times 10^{-16}~\mathrm{s}^{-1}$, are consistent and in the same order as our results. We also find a similar agreement with the adapted results of Porras et al.~(\cite{Porras+2014}) with a range of $\zeta_p$-values equal to $(2.2-20.6) \times 10^{-16}~\mathrm{s}^{-1}$, though the average value is about 50 percent higher ($12.1 \times 10^{-16}~\mathrm{s}^{-1}$). The exceptions are three interstellar velocity components in the sightlines towards HD\,149404, HD\,154368, and HD\,183143. In those cases, the OH$^+$ lines are very weak (not even the CH$^+$ counterpart\footnote{This is based on studies by, e.g., Kre\l{}owski et al.~(\cite{Krelowski+2010}) and Porras et al.~(\cite{Porras+2014}) which suggest that OH$^+$ and CH$^+$ are associated with each other.} is detected), suggesting that these extreme values should be viewed with caution. Also, care must be taken in directly comparing the results obtained from individual velocity components and from total sightline measurements (as in this work). Despite the seemingly similar ranges, it should be noted that these comparisons should not be taken at face value since individual targets are likely to be in quite unique physical environments, as shown by the differences in $\zeta_p$ of up to more than an order of magnitude. A thorough statistical treatment of the data may help distinguish any difference in the distribution of the sightlines.
        
    As for the 28 targets with (weak) single-OH$^+$ line detection (Appendix~\ref{asec:moretargets}), we get a range of $(1.3-9.4) \times 10^{-16}~\mathrm{s}^{-1}$ which overlaps with or is close to our results for the 10 main targets but does so on the lower side of the range. The disparity becomes more clear when comparing weighted averages; the single-line targets have an average $\zeta_p$-value of $3.0(3) \times 10^{-16}~\mathrm{s}^{-1}$ which is about three times lower than what we have for the multi-line targets. Looking at the OH$^+$ column densities, we see that the results for the 10 selected lines-of-sight with multiple transitions (Fig.~\ref{fig:grid}) are systematically higher ($(1.3-8.3) \times 10^{13}~\mathrm{cm}^{-2}$) than those derived for the 28 lines-of-sight for which only one transition (line 1) is observed ($(0.5-2.4) \times 10^{13}~\mathrm{cm}^{-2}$). This difference cannot be attributed by having a [more accurate] multiple line fit, as focusing only on the transition with the largest oscillator strength (line 1) in these lines-of-sight results in comparable (though less accurate) $N$(OH$^+$)-values. The selection of targets with multiple OH$^+$ transitions comes with a bias, namely, that the $N$(OH$^+$)-values measured for those environments are larger. The $N$(OH$^+$) abundances clearly span a range of values somewhat more than an order of magnitude and this results in a range of $\zeta_p$-values as well; similarly, variations in $N$(H\textsubscript{tot}) also affect the derived $\zeta_p$-values. When the 10 main and 28 additional targets are taken together, we get an average $\zeta_p$-value of $5.1(3) \times 10^{-16}~\mathrm{s}^{-1}$.
    
    \begin{figure}
        \centering
        \includegraphics[width=0.9\hsize]{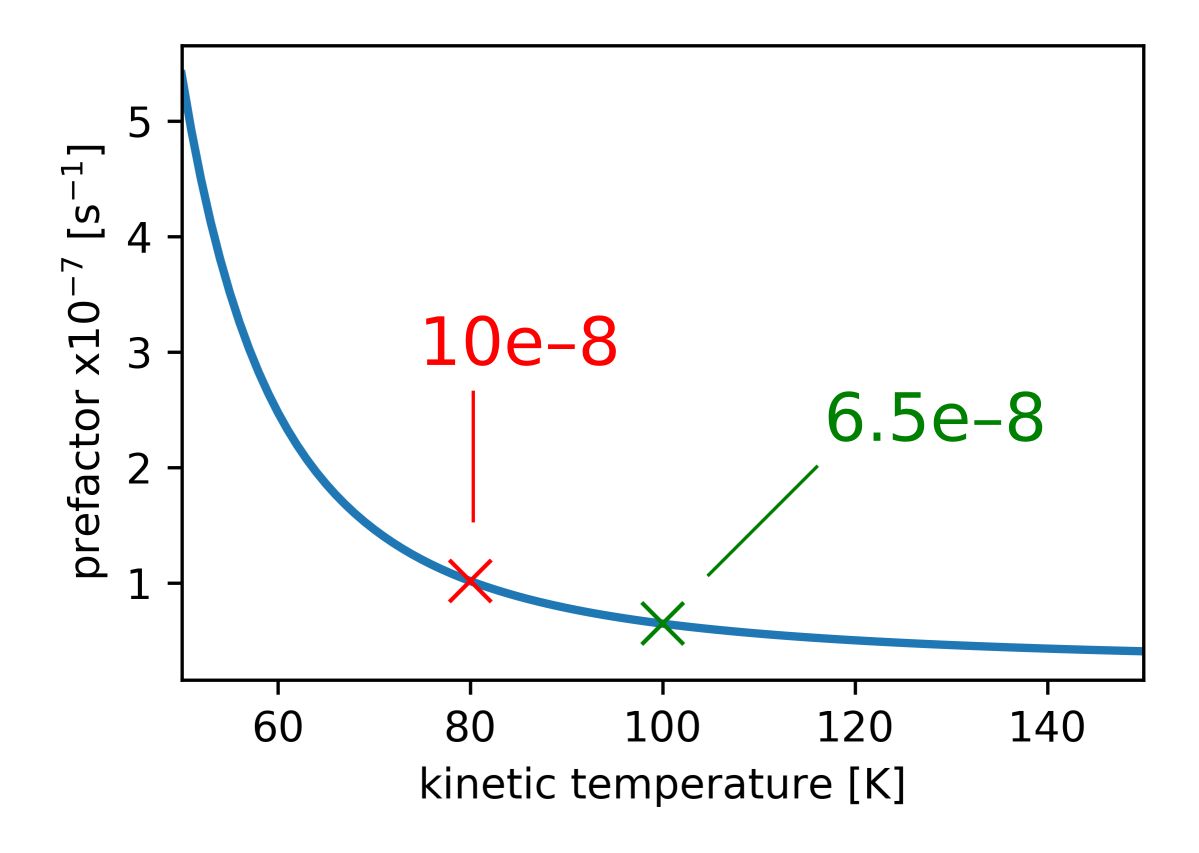}
        \caption{\label{fig:Tdependence}A graph showing the effect of the temperature dependence of the reaction rate constants to the resulting prefactor in Eq.~\ref{eq:CRIR}. The prefactor for $T = 80$~K and $T = 100$~K are highlighted in red and green, respectively.
        }
    \end{figure}
    
    We can also compare our work with other cosmic ray ionization rate investigations using other methods and tracers. Noting that $\zeta_p$ is approximately related to the cosmic ray ionization rates of atomic and molecular hydrogen by $\zeta_p = \zeta_{\mathrm{H}}/1.5 = \zeta_{\mathrm{H}_2}/2.3$ (Glassgold \& Langer~\cite{Glassgold&Langer1974}), we find that our results are generally higher. These comparisons include detections of OH$^+$ and H$_2$O$^+$ in the sub-mm region (Neufeld et al.~\cite{Neufeld+2010}; Indriolo et al.~\cite{Indriolo+2012}) which give $\zeta_p \sim (0.4-3.0) \times 10^{-16}~\mathrm{s}^{-1}$ and H$_3^+$ in the infrared (Le Petit et al.~\cite{LePetit+2004}; Indriolo et al.~\cite{Indriolo+2007}; Indriolo \& McCall~\cite{Indriolo&McCall2012}; Indriolo et al.~\cite{Indriolo+2012,Indriolo+2015}; Neufeld \& Wolfire~\cite{Neufeld&Wolfire2017}) which give $\zeta_p \sim (0.5-4.6) \times 10^{-16}~\mathrm{s}^{-1}$. Although McCall et al.~(\cite{McCall+2003}) had derived a high $\zeta_p$-value of $12 \times 10^{-16}~\mathrm{s}^{-1}$ using H$_3^+$ observations along the sightline towards $\zeta$~Persei, this was subsequently updated to a much lower value of $2.5 \times 10^{-16}~\mathrm{s}^{-1}$ by LePetit et al.~(\cite{LePetit+2004}) with a more detailed photodissociation region (PDR) cloud model. 
    
    Comparing results obtained using the same tracer can shed some light on the possible environmental differences in the various sightlines studied thus far. Another useful exercise is to see how different tracers of the cosmic ray ionization rate compare for the same sightline. This will indicate how well our existing models for different tracers are able to describe the mechanism behind the processes taking place in these environments. In Table~\ref{atab:OH+H3+}, we list sightlines from our EDIBLES data set (as well as from Kre\l{}owski et al.~\cite{Krelowski+2010} and Porras et al.~\cite{Porras+2014}) where we have OH$^+$ data (or upper limits) with corresponding H$_3^+$ data (or upper limits) from the work of Indriolo \& McCall~(\cite{Indriolo&McCall2012}) and Albertsson et al.~(\cite{Albertsson+2014}). Note that all of the targets listed have a single OH$^+$ absorption line measured apart from HD~41117 where we have detected 7 lines. For this target, the $\zeta_p$-value we get from our work is about five times higher than the value obtained by Albertsson et al. Other targets with both OH$^+$ and H$_3^+$ detection show more or less the same, overlapping values (HD~24398, HD~110432, HD~154368, HD~169454). For HD~183143, a measurent by Porras et al. and by Indriolo \& McCall corresponds to the same velocity component ($v_\odot = -10$~km\,s$^{-1}$ and $v_\mathrm{LSR} = 7$~km\,s$^{-1}$, respectively), but the derived $\zeta_p$-values differ by about five times. As for upper limits, care should be taken in comparing them. It should be kept in mind that the cosmic ray ionization rate values deduced will rely on specific assumptions regarding, e.g., density and temperature. However, the general trend that the $\zeta_p$-values derived from OH$^+$ observations is larger than that derived from H$_3^+$ may reflect an actual decrease of this quantity from the edge of these molecular clouds towards the center which may be exhibited thanks to the spatial stratification of the OH$^+$ and H$_3^+$ molecular ions. Such a possibility was raised independently by Rimmer et al.~(\cite{Rimmer+2012}) in order to understand the presence of carbon chains in the illuminated part of the Horsehead nebula. A recent theoretical study on the penetration of cosmic rays in diffuse clouds by Phan et al.~(\cite{Phan+2018}) also points out to that possibility. The results we obtained for diffuse clouds follow the general trend of values being an order of magnitude larger than the cosmic ray ionization rates found in dense molecular clouds (van der Tak \& van Dishoeck~\cite{vanderTak&vanDishoeck2000}; Kulesa~\cite{Kulesa2002}).
    
    With all of these comparisons it is important to realize how, in our formulation, the prefactor in Eq.~\ref{eq:CRIR} is influenced by a variety of parameters. One of these is the assumption made regarding the relative fractional abundances of hydrogen, i.e., $x$(H) = 10$x$(H$_2$). As can be seen in Table~\ref{tab:targets}, the relative hydrogen column densities do not necessarily follow this relationship, and most of them fall short on a factor 10. Thus, if we use the actual measured column densities of H~\textsc{i} and H\textsubscript{2} in our formulation, this would yield a unique prefactor in Eq.~\ref{eq:CRIR} for every sightline. Other factors such as the O and the PAH$^{(-)}$ abundances may also play crucial roles in the resulting $\zeta_p$-values. Clearly, there exists a need for a thorough investigation on how these parameters, as well as the properties of the individual sightlines, dictate the numbers that we get.    
    
    Another factor worth considering is the temperature dependence of the prefactor, mainly driven by the H$^+$ + O charge-exchange rate coefficient $k_1$. This charge-exchange takes place with atomic oxygen in its $J = 2$ ground level and has an exp($-227/T$) temperature dependence, corresponding to the endothermicity of the reaction. It can be seen from the abundance equation for OH$^+$ (Sec.~\ref{sec:CRIR}) that $k_1$ is the only $T$-dependent factor in the numerator of $x$(OH$^+$) and dominates the denominator at higher $T$ which causes the steep rise in the prefactor as $T$ falls below 60~K and the gradual decline as $T$ rises above 100~K (with other reactions having weaker $T$-dependence). The overall $T$-dependence is shown in Fig.~\ref{fig:Tdependence} -- a change in $T$ from 80~K to 100~K can give a difference in the prefactor by about 1.5 times. (For our case, we take a temperature of $T = 100$~K as the typical temperature in diffuse clouds.) Four of our sightlines have data for $T_{01}$ (Rachford et al.~\cite{Rachford+2002},~\cite{Rachford+2009}; Sheffer et al.~\cite{Sheffer+2008}) which range from 59~K for HD~41117 and 101~K for HD~185418, and these correspond to cosmic ray ionization rates which may vary by about a factor of four. Moreover, reactions between molecular hydrogen and O$^+$/OH$^+$ have also been recently studied in ion trap experiments at low temperatures (Kovalenko et al.~\cite{Kovalenko+2018}; Tran et al.~\cite{Tran+2018}). The derived rate coefficients corresponding to $k_3$ and $k_4$ are within the same order of magnitude as with previous values, which give differences in the calculated $\zeta_p$ by at most 20 percent. (We keep the values of $k_3$ and $k_4$ used by other authors for the purpose of a consistent comparison with other near-UV studies.) Given these existing dependencies, care is needed in using specific cosmic ray ionization rates.
    
    Searching for these near-UV OH$^+$ transitions in other galaxies is also promising. There have already been a number of extragalactic detections of OH$^+$, which include studies by van der Werf et al.~(\cite{vanderWerf+2010}), Gonz\'ales-Alfonso et al.~(\cite{Gonzales-Alfonso+2013}),  Riechers et al.~(\cite{Riechers+2013}), and recently by Muller et al.~(\cite{Muller+2016}) who have found slightly higher values of $\zeta_p$ for the $z = 0.89$ absorber PKS 1830-211 measured within a similar galactocentric radius as in studies of the Milky Way (Indriolo et al.~\cite{Indriolo+2015}); they attributed this to the higher star formation rate in the former. They got values of $\zeta_p \sim 130 \times 10^{-16}$ and $20 \times 10^{-16}~\mathrm{s}^{-1}$, along sightlines located at $\sim$2~kpc and $\sim 4$~kpc to either side of the galactic center, respectively. 
    Recently, Indriolo et al.~(\cite{Indriolo+2018}) have also reported $\zeta_p$-values ranging from the high $\sim$ 10$^{-17}$ to 10$^{-15}$ towards the $z\sim2.3$ lensed galaxies SMM J2135-0102 and SDP 17b from observations of both OH$^+$ and H$_2$O$^+$.
    All of these studies have so far looked at the sub-mm transitions of OH$^+$, and thus, having complementary observations in the near-UV and in the optical for these [high-redshift] extragalactic (albeit faint) targets would help us in probing variations of the cosmic ray ionization rate over cosmic timescales.

\section{Conclusions}

    In this contribution, we have determined cosmic ray ionization rates along 10 diffuse interstellar sightlines through the measurement of OH$^+$ abundances which are constrained better with the detection of more near-UV OH$^+$ electronic transitions. The explicit incorporation of proton recombinations on PAHs increases the historically used prefactor for the $N$(OH$^+$)/$N$(H\textsubscript{tot}) ratio and results in larger cosmic ray ionization rates. We obtain a range of $\zeta_p$-values equal to $(3.9-16.4) \times 10^{-16}~\mathrm{s}^{-1}$, which is generally much higher than what was derived in previous studies from detections of interstellar OH$^+$ in the far-infrared / sub-millimeter-wave regions but is comparable to measurements in the near-ultraviolet using a reformulated abundance equation for interstellar OH$^+$, as introduced here. An additional constraint on the physical conditions prevailing in these diffuse lines-of-sight, and on the derived primary cosmic ray ionization rate, could be obtained through the detection of H$_2$O$^+$ absorption transitions which occur in the visible (Lew~\cite{Lew1976}; Gredel et al.~\cite{Gredel+2001}). This ion has been detected in the ISM in the infrared by \emph{Herschel} (e.g., Ossenkopf et al.~\cite{Ossenkopf+2010}) but not yet in the optical. H$_2$O$^+$ is indeed formed directly through the OH$^+$~+~H$_2$ reaction and its destruction results from dissociative recombination by electrons and a further reaction with H$_2$. These signatures will be searched in the EDIBLES spectra. Finally, it will be interesting to investigate whether the (non)detection of OH$^+$ can be linked to the (non)appearance of specific DIBs, as currently is being investigated for EDIBLES data linking selected DIBs to C$_2$ (Elyajouri et al.~\cite{Elyajouri2018}).

\begin{acknowledgements}
    This work is based on observations obtained at the European Organization for Astronomical Research in the Southern Hemisphere under ESO programs 194.C-0833 and 266.D-5655. XLB thanks Edcel Salumbides for fruitful discussions and the whole EDIBLES team for their support and for making the reduced UVES data available. JC and AF acknowledge support from an NSERC Discovery Grant and a Western Accelerator Award. HL acknowledges support through NOVA and NWO.
\end{acknowledgements}

\balance

\newpage

\begin{appendix}

\onecolumn

\section{EDIBLES targets with very weak OH$^+$ absorptions}\label{asec:moretargets}
    
    Table~\ref{atab:moretargets} is a comprehensive list of additional EDIBLES sightlines that show some weak OH$^+$ absorption that can be discerned through visual inspection, but are excluded in the present analysis.
    
    \begin{table}[h!]
        \centering
        \caption{Estimated cosmic ray ionization rates for EDIBLES targets with a single OH$^+$ absorption ($\lambda3584$).}
        \label{atab:moretargets}
        \resizebox{\textwidth}{!}{%
        \begin{tabular}{@{}lcccc|lcccc@{}}
        \toprule
        \midrule
        \multicolumn{1}{c}{Identifier} & \begin{tabular}[c]{@{}c@{}}$W_{3584}$\\ {[}m$\AA${]}\end{tabular} & \begin{tabular}[c]{@{}c@{}}$N$(OH$^+$)\\ $\times 10^{13}$ {[}cm$^{-2}${]}\end{tabular} & \begin{tabular}[c]{@{}c@{}}$N$(H$_\mathrm{tot}$)\\ $\times 10^{21}$ {[}cm$^{-2}${]}\end{tabular} & \begin{tabular}[c]{@{}c@{}}$\zeta_p$\\ $\times 10^{-16}$ {[}s$^{-1}${]}\end{tabular} & \multicolumn{1}{c}{Identifier} & \begin{tabular}[c]{@{}c@{}}$W_{3584}$\\ {[}m$\AA${]}\end{tabular} & \begin{tabular}[c]{@{}c@{}}$N$(OH$^+$)\\ $\times 10^{13}$ {[}cm$^{-2}${]}\end{tabular} & \begin{tabular}[c]{@{}c@{}}$N$(H$_\mathrm{tot}$)\\ $\times 10^{21}$ {[}cm$^{-2}${]}\end{tabular} & \begin{tabular}[c]{@{}c@{}}$\zeta_p$\\ $\times 10^{-16}$ {[}s$^{-1}${]}\end{tabular} \\ \midrule
        HD 22951    & 0.4(3)        & 0.7(5)        & 1.7   & 2.7(1.8)          & HD 152408 & 1.4(7)        & 2.4(1.2)          & 2.3       & 6.7(3.3)      \\
        HD 23180    & 0.3(3)        & 0.5(4)        & 1.6   & 2.2(1.7)          & HD 152424 & 1.3(7)        & 2.1(1.1)          & 3.8*      & 3.6(1.8)    \\
        HD 24398    & 0.3(3)        & 0.5(4)        & 1.6   & 2.1(1.8)          & HD 154043 & 0.9(6)        & 1.5(9)        & 4.6*      & 2.2(1.3)    \\
        HD 37903  &\textit{0.5(5)} &\textit{0.9(8)} & 2.7   & \textit{2.0(1.9)} & HD 155806 & 0.7(3)        & 1.1(4)        & 1.4       & 5.4(2.0)      \\
        HD 75309    & 0.9(6)          & 1.6(1.0)          & 1.5   & 6.9(4.4)          & HD 166937 & 0.8(2)        & 1.4(3)        & 1.3*      & 7.1(1.5)      \\
        HD 111934   & 0.7(6)        & 1.2(9)        & 2.0*  & 4.0(3.1)          & HD 167264 & 0.7(3)        & 1.1(5)        & 1.8       & 4.1(1.9)    \\
        HD 113904   & 0.5(3)        & 0.8(6)        & 1.3   & 3.8(2.7)          & HD 167838 & 0.6(6)        & 0.9(1.0)          & 4.1*      & 1.5(1.7)    \\
        HD 122879 &\textit{0.4(4)} &\textit{0.6(6)} & 2.2   & \textit{1.9(1.8)} & HD 169454 & 0.8(9)        & 1.4(1.4)          & 6.9*      & 1.3(1.4)    \\
        HD 145502 &\textit{0.5(4)} &\textit{0.8(7)} & 1.3   & \textit{3.8(3.3)} & HD 170740 & 0.7(3)        & 1.1(5)        & 2.5       & 2.8(1.2)    \\
        HD 148937   & 0.8(6)        & 1.4(1.0)          & 5.0   & 1.8(1.2)          & HD 171957 &\textit{0.8(4)}&\textit{1.4(6)}& 1.6*      &\textit{5.8(2.5)}\\
        HD 149038   & 1.3(1.1)          & 2.3(1.8)          & 1.6   & 9.4(7.3)          & HD 172694 & 1.4(5)        & 2.3(8)        & 2.0*      & 7.6(2.5)      \\
        HD 149404   & 0.7(3)        & 1.2(5)        & 3.9*  & 1.9(8)          & HD 180554 & 0.4(2)        & 0.7(4)        & $\cdots$  & $\cdots$  \\
        HD 151804   & 0.9(5)        & 1.6(8)        & 1.6   & 6.4(3.1)          & HD 184915 & 0.3(2)        & 0.6(3)        & 1.1       & 3.3(1.9)      \\
        HD 152248   & 0.6(4)        & 1.0(7) &$\geq$\,1.7   &$\leq$\,3.9      & HD 303308 & 0.9(8)          & 1.7(1.3)          & 3.0       & 3.6(2.8)      \\ \bottomrule
        \end{tabular}%
        }
        \tablefoot{For weak absorption features with multiple components, an integrated area over $\sim1~\AA$ centered at $\lambda3584$ is reported; these values are indicated by the italicized numbers. The $N$(H\textsubscript{tot})-values are derived from Cox et al.~(\cite{Cox+2017}); the ones with an asterisk (*) are derived (as described in Sec.~\ref{sec:reddening&extinction}) from the $E_{B-V}$-values taken from the same reference. Numbers enclosed in parentheses denote the uncertainty of the last digit(s); e.g., 1.2(3) $\equiv$ 1.2~$\pm$~0.3, whereas 4.3(2.1) $\equiv$ 4.3~$\pm$~2.1.
        }
    \end{table}

\section{Compilation of estimates of $\zeta_p$ derived from OH$^+$ detections in the near-UV}\label{asec:literature}
    
    \vskip -0.5cm
    
    \begin{table}[h!]
        \caption{Literature values for $N$(OH$^+$) together with the [estimated] $N$(H\textsubscript{tot}) column densities and the corresponding $\zeta_p$, both the values reported originally and the values adapted here to follow Eq.~\ref{eq:CRIR}.}
        \label{atab:literature}
        \centering
        \begin{tabular}{l c c c c}\toprule\midrule
        \multirow{2}{*}{Target}      & scaled $N$(OH$^+$)            & $N$(H\textsubscript{tot})            & original $\zeta_p$      & adapted $\zeta_p$      \\
                    & $\times10^{13}$ [cm$^{-2}$]   & $\times10^{21}$ [cm$^{-2}$]   & $\times10^{-16}$ [s$^{-1}$]  & $\times10^{-16}$ [s$^{-1}$] \\ \midrule
        Porras et al.~\cite{Porras+2014}\\
        \cmidrule(lr){0-0}
        BD-14 5037  & 0.58                   & 0.22          & 1.6       & 17.3         \\
                    & 1.9                    & 2.3           & 0.5       & 5.4         \\
        HD 149404   & 0.41                   & 1.2           & 0.2       & 2.2         \\
                    & 0.87\tablefootmark{\textcolor{blue}{a}}  & 0.095\tablefootmark{\textcolor{blue}{a}}          & 5.5\tablefootmark{\textcolor{blue}{a}}       & 59.5\tablefootmark{\textcolor{blue}{a}}         \\
                    & 1.1                    & 0.54          & 1.2       & 13.0         \\
                    & 0.65                   & 0.30          & 1.3       & 14.1         \\
        HD 154368   & 0.37\tablefootmark{\textcolor{blue}{a}}  & 0.067\tablefootmark{\textcolor{blue}{a}}         & 3.3\tablefootmark{\textcolor{blue}{a}}       & 35.7\tablefootmark{\textcolor{blue}{a}}         \\
                    & 0.91                   & 1.1           & 0.5       & 5.4         \\
                    & 0.35                   & 0.16          & 1.3       & 14.1         \\
        HD 183143   & 2.4                    & 0.75          & 1.9       & 20.6         \\
                    & 0.61\tablefootmark{\textcolor{blue}{a}}  & 0.060\tablefootmark{\textcolor{blue}{a}}         & 6.1\tablefootmark{\textcolor{blue}{a}}       & 66.0\tablefootmark{\textcolor{blue}{a}}         \\
                    & 0.67                   & 0.25          & 1.6       & 17.3         \\
        \\
        Zhao et al.~\cite{Zhao+2015}\\
        \cmidrule(lr){0-0}
        CD-32 4348  & 6.3(3)        & 5.7        & 0.8         & 7.2(3)         \\
        HD 63804    & 7.7(3)        & 4.5        & 1.2         & 11.1(4)        \\
        HD 78344    & 4.0(4)        & 4.0        & 0.8         & 6.6(7)         \\
        HD 80077    & 4.2(6)        & 3.7        & 0.9         & 7.3(1.0)       \\
        \bottomrule
        \end{tabular}
        \tablefoot{The OH$^+$ column densities reported by Porras et al.~(\cite{Porras+2014}) were derived from a single electronic transition ($\lambda$3584) while those of Zhao et al.~(\cite{Zhao+2015}) result from a line fit through multiple OH$^+$ absorption lines, as in this work. These values have been scaled according to the recently updated line oscillator strengths provided by Hodges et al.~(\cite{Hodges+2018}). Numbers enclosed in parentheses denote the uncertainty of the last digit(s); e.g., 1.2(3) $\equiv$ 1.2~$\pm$~0.3, whereas 4.3(2.1) $\equiv$ 4.3~$\pm$~2.1.
        \\
        \tablefootmark{a} Values reported for these components correspond to very weak OH$^+$ absorption components with no corresponding CH$^+$ detections. The associated $N$(H\textsubscript{tot}) is likely underestimated leading to overestimated values for $\zeta_p$.
        }
    \end{table}

    \newpage

\section{Linear regression results for \emph{N}(OH$^+$)}\label{asec:N(OH+)}
    
    \begin{figure*}[h!]
        \centering
        \includegraphics[width=0.95\textwidth]{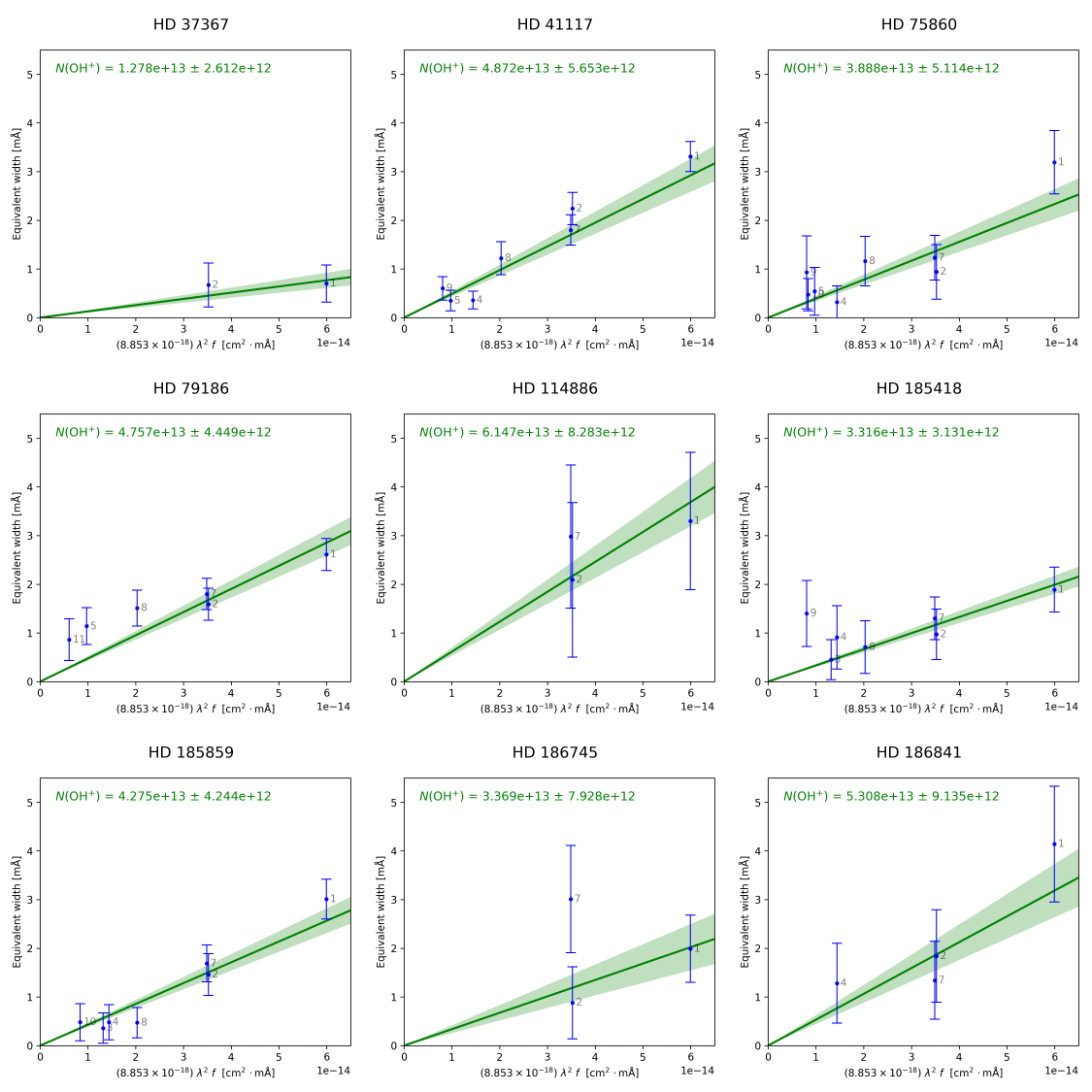}
        \caption{\label{afig:N(OH+)}Plotted above are the weighted linear fits for deriving $N$(OH$^+$) for all targets except HD~80558 (Fig.~\ref{fig:N(OH+)}). A couple of outliers can be noticed for some of the sightlines as in the cases for HD~185418 and HD~186745 which could well be caused by poor SNR (see Fig.~\ref{fig:grid}) and/or possible contamination with some weak absorption features. This behavior is also seen in the measurements for the weaker OH$^+$ lines in HD~79186. Setting the intercept to zero helps constrain the effect of these biases.
        }
    \end{figure*}

\newpage

\section{Compilation of $\zeta_p$-values derived from OH$^+$ and H$\mathsf{_3}^+$ studies}\label{asec:OH+H3+}

    \begin{table}[h!]
        \centering
        \caption{A comparison of primary cosmic ray ionization rates $\zeta_p$ among targets which have detections for both OH$^+$ (this work; Kre\l{}owski et al.~\cite{Krelowski+2010}; Porras et al.~\cite{Porras+2014}) and H${_3}^+$ (Indriolo \& McCall~\cite{Indriolo&McCall2012}; Albertsson et al.~\cite{Albertsson+2014}).}
        \label{atab:OH+H3+}
        \begin{tabular}{@{}llcccccc@{}}
        \toprule
        \midrule
        \multirow{2}{*}{Identifier} &  & \multicolumn{3}{c}{OH$^+$ $\zeta_p$ $\times10^{-16}$ s$^{-1}$} &  & \multicolumn{2}{c}{H${_3}^+$ $\zeta_p$ $\times10^{-16}$ s$^{-1}$} \\ \cmidrule(lr){3-5} \cmidrule(l){7-8} 
                                    &  & this work    & Kre\l{}owski et al.~\cite{Krelowski+2010}    & Porras et al.~\cite{Porras+2014}   &  & Indriolo \& McCall~\cite{Indriolo&McCall2012} & Albertsson et al.~\cite{Albertsson+2014}\\ \midrule
        HD 22951   &  & 2.7(1.8)        & $\cdots$ & $\cdots$                                       &  & $\leq$\,1.2      & $\cdots$    \\
        HD 23180   &  & 2.2(1.7)       & $\cdots$ & $\cdots$                                       &  & $\leq$\,1.8       & $\cdots$    \\
        HD 24398   &  & 2.1(1.8)        & $\cdots$ & $\cdots$                                       &  & 2.4(1.4)       & $\cdots$    \\
        HD 27778   &  & $\leq$\,9.8     & $\cdots$ & $\cdots$                                       &  & $\leq$\,4.5 & 2.3(3)      \\
        HD 41117   &  & 10.3(8)     & $\cdots$ & $\cdots$                                       &  & $\leq$\,6.0 & 2.3(7)      \\
        HD 43384   &  & $\leq$\,8.6     & $\cdots$ & $\cdots$                                       &  & $\cdots$   & 1.1(3)      \\
        HD 110432  &  & $\cdots$    & 2.7(1.4)     & $\cdots$                                       &  & 1.7(9)     & $\cdots$    \\
        HD 147888   &  & $\leq$\,2.2       & $\cdots$ & $\cdots$ &  & $\leq$\,20.1       & $\cdots$    \\
        HD 147889   &  & $\leq$\,15.7       & $\cdots$ & $\cdots$ &  & $\leq$\,0.8       & $\cdots$    \\
        HD 149038   &  & 9.4(7.3)       & $\cdots$ & $\cdots$ &  & $\leq$\,2.4       & $\cdots$    \\
        HD 149404   &  & 1.9(8)       & 6.1(3.1) & 2.2, 59.5,\tablefootmark{\textcolor{blue}{a}} 13.0, 14.1                                       &  & $\leq$\,1.9       & $\cdots$    \\
        HD 149757   &  & $\leq$\,5.3       & $\cdots$ & $\cdots$ &  & $\leq$\,0.8       & $\cdots$    \\
        HD 154368  &  & $\cdots$    & 3.4(2.0)     & 35.7,\tablefootmark{\textcolor{blue}{a}} 5.4, 14.1 &  & 1.8(1.1)       & $\cdots$    \\
        HD 169454  &  &1.3(1.4)      & $\cdots$ & $\cdots$                                       &  & 1.1(8)     & $\cdots$    \\
        HD 183143  &  & $\cdots$    & $\cdots$ & 20.6*, 66.0,\tablefootmark{\textcolor{blue}{a}} 17.3  &  & 4.6(3.6)*, 3.4(2.6) & $\cdots$    \\
        BD-14\,5037  &  & $\cdots$    & $\cdots$ & 17.3, 5.4  &  & $\leq$\,0.3 & $\cdots$    \\
        \bottomrule
        \end{tabular}
        \tablefoot{The HD~149404 data from Porras et al.~(\cite{Porras+2014}) come from four velocity components, while the  HD~154368 and HD~183143 data come from three. The HD~183143 data from Indriolo \& McCall~(\cite{Indriolo&McCall2012}) come from two velocity components, one of which, at $v_\mathrm{LSR} = 7$~km\,s$^{-1}$, corresponds to a component measured by Porras et al. at $v_\odot = -10$~km\,s$^{-1}$ (each component marked with an asterisk). Upper limits reported by Indriolo \& McCall are expressed here as 5$\sigma$ results. Values reported in the last column have been computed with the same assumptions as in Indriolo \& McCall~(\cite{Indriolo&McCall2012}).
        Numbers enclosed in parentheses denote the uncertainty of the last digit(s); e.g., 1.2(3) $\equiv$ 1.2~$\pm$~0.3, whereas 4.3(2.1) $\equiv$ 4.3~$\pm$~2.1.
        \\
        \tablefootmark{a} Values reported for these components correspond to very weak OH$^+$ absorption components with no corresponding CH$^+$ detections. The associated $N$(H\textsubscript{tot}) is likely underestimated leading to overestimated values for $\zeta_p$.
        }
    \end{table}

\end{appendix}

\end{document}